\documentclass{aa}

\usepackage{amsmath,amssymb,textcomp}
\usepackage{graphicx}
\usepackage[varg]{txfonts}
\usepackage{enumitem}

\usepackage{natbib}

\usepackage{import}

\usepackage{hyperref}
\hypersetup{pdfborder={0 0 0}}
\hypersetup{
     colorlinks = true, 
     linkcolor = blue,
     anchorcolor = black,
     citecolor = blue, 
     filecolor = black, 
     urlcolor = blue
}

\usepackage{multirow}

\makeatletter
\def\hlinewd#1{%
\noalign{\ifnum0=`}\fi\hrule \@height #1 %
\futurelet\reserved@a\@xhline}
\makeatother
\newcommand{\dif}{\mathrm{d}}
\newcommand{\idem}{~~\textacutedbl}
\newcommand{\tot}{\mathrm{tot}}
\newcommand{\cs}{\mathrm{cs}}
\newcommand{\bg}{\mathrm{bg}}
\newcommand{\out}{\mathrm{out}}
\newcommand{\ins}{\mathrm{in}}
\newcommand{\lec}{\mathrm{e}}
\renewcommand{\ion}{\mathrm{i}}

\newcommand{\pe}{\mathrm{pe}}
\newcommand{\ce}{\mathrm{ce}}

\newcommand{\wci}{\omega_\mathrm{ci}}
\newcommand{\wce}{\omega_\mathrm{ce}}
\newcommand{\wpe}{\omega_\mathrm{pe}}
\newcommand{\wpi}{\omega_\mathrm{pi}}

\def\dintcl{\int\limits_{\mathrm{cell}}\kern-0.7em\int\kern-1.4em\bigcirc\kern.7em}
\renewcommand{\b}{\textbf}

\usepackage{color}
\definecolor{orange}{rgb}{0.6,0.1,0}
\definecolor{darkblue}{rgb}{0.,0.,0.4}

\renewcommand{\note}[1]{}
\newcommand{\modif}{\textcolor{orange}}

\begin{document}

\title{The energetics of relativistic magnetic reconnection: ion-electron repartition and particle distribution hardness}

\author{Micka\"el Melzani\inst{1}         \and
        Rolf Walder\inst{1}               \and
        Doris Folini\inst{1}              \and 
        Christophe Winisdoerffer\inst{1}  \and 
        Jean M. Favre\inst{2}
        }

\institute{\'{E}cole Normale Sup\'{e}rieure, Lyon, CRAL, UMR CNRS 5574, 
           Universit\'{e} de Lyon, France\\
           \hspace{0.3cm} E-mail: mickael.melzani@ens-lyon.fr
           \and
           CSCS Lugano, Switzerland
}

\offprints{M. Melzani}

\date{Received ... ; accepted ...}

\authorrunning{M. Melzani et al.}
\titlerunning{The energetics of relativistic magnetic reconnection}


\abstract{%
Collisionless magnetic reconnection is a prime candidate to account for flare-like or steady emission, outflow launching, 
or plasma heating,
in a variety of high-energy astrophysical objects, including ones with relativistic ion-electron plasmas.
But the fate of the initial magnetic energy in a reconnection event remains poorly known:
what is the amount given to kinetic energy, the ion/electron repartition, and the hardness of the particle distributions? 
We explore these questions with 2D particle-in-cell simulations of ion-electron plasmas.
We find that $45$ to $75\%$ of the total initial magnetic energy ends up in kinetic energy, 
this fraction increasing with the inflow magnetization.
Depending on the guide field strength, ions get from 30 to $60\%$ of the total kinetic energy.
Particles can be separated into two populations that only weakly mix: 
(i) particles initially in the current sheet, heated by its initial tearing and subsequent contraction of the islands; 
and (ii) particles from the background plasma that primarily gain energy via the reconnection electric field when passing near the X-point.
Particles (ii) tend to form a power-law with an index $p=-\dif\log n(\gamma)/\dif\log\gamma$, 
that depends mostly on the inflow Alfv\'en speed $V_\mathrm{A}$ and magnetization $\sigma_s$ of species $s$, 
with for electrons $p=5$ to 1.2 for increasing $\sigma_\lec$.
The highest particle Lorentz factor, for ions or electrons, 
increases roughly linearly with time for all the relativistic simulations.
This is faster, and the spectra can be harder, than for collisionless shock acceleration.
We discuss applications to microquasar and AGN coronae, to extragalactic jets, and to radio lobes.
We point out situations where effects such as Compton drag or pair creation are important.
}%


\maketitle

\begin{table*}[tbp]
\caption{\label{tab:param_magnetization}
Parameters of the inflow (or background) plasma, and resulting power-law index $p$, 
sorted in order of increasing magnetization (except for the last line).
The parameters of the current sheet, for each value $(\wce/\wpe,\,m_\ion/m_\lec)$, are listed in Table~\ref{tab:param_physical_tearing}.
The background plasma $\beta_s = n_sT_s/(B^2/2\mu_0) = 2\Theta_s/\sigma_{s}^\mathrm{cold}(B)$ includes the guide field.
The magnetization $\sigma^\mathrm{hot}_s$ is defined by Eq.~\ref{equ:acc_sigma_s}, and 
$\sigma_{\ion+\lec}$ is the total magnetization (Eq.~\ref{equ:def_Alfven_speed}).
The Alfv\'en speed ${V^\mathrm{R}_\mathrm{A,in}}$ is defined by Eq.~\ref{equ:sigma_tot}.
When there is a guide field, the value displayed is $V^\mathrm{R}_\mathrm{A,in}\cos\theta$ (Sect.~\ref{sec:the_simulations}).
The index of the power-law component of the background population (when there is one) is $p$.
}
\centering
\begin{tabular}{c|c|c||c|c||c|c|c|c||c}
 $\wce/\wpe$ & $n_\mathrm{bg}/n_\cs(0)$ & $B_\mathrm{G}/B_0$ & $\sigma_{\ion+\lec}$ & $V^\mathrm{R}_\mathrm{A,in}/c$ & & $T_{\bg,s}$ (K) & $\beta_s$ & $\sigma^\mathrm{hot}_s(B_\mathrm{rec})$ & $p = -\dif \log n(\gamma) / \dif \log\gamma$ \\
\hline
 1 & 0.1 & 0   & 0.38 & 0.53 &  ion  & $1.5\times10^7$ & $5\times10^{-4}$   & 0.4  & no power-law \\
 {\small($m_\ion=25m_\lec$)} & & & & & elec.& \idem    & \idem              & 9.9  & 4-5 \\
\hline
 3 & 0.31& 0   & 1.11 & 0.73 &  ion  & $2\times10^8$   & $2.5\times10^{-3}$ & 1.2  & 8 \\
 {\small($m_\ion=25m_\lec$)} & & & & &  elec.& \idem   & \idem              & 27   & 2.2-2.6 \\
\hline
 3 & 0.1 & 0   & 3.26 & 0.88 &  ion  & $2\times10^8$   & $7.5\times10^{-4}$ &  3.6 & 6.5 \\
 {\small($m_\ion=25m_\lec$)} & & & & &  elec.& $3\times10^9$   & $1.1\times10^{-2}$ &  35  & 2.8 \\
\hline
 3 & 0.1 & 0   & 3.46 & 0.88 &  ion  & $1.5\times10^7$ & $5.6\times10^{-5}$ &  3.6 & not investigated \\
 {\small($m_\ion=25m_\lec$)} & & & & &  elec.& \idem   & \idem              &  89  & \idem \\
\hline
 3 & 0.1 & 0   & 3.45 & 0.88 &  ion  & $2\times10^8$   & $7.5\times10^{-4}$ &  3.6 & 5.8 \\
 {\small($m_\ion=25m_\lec$)} & & & & &  elec.& \idem   & \idem              &  83  & 1.5-2 \\
   & \multicolumn{8}{c||}{Identical as above, but larger box ($888\times1138\,d_\lec$ instead of $455\times683\,d_\lec$) and duration}  & 4.8 \\
   & \multicolumn{8}{c||}{}  & 1.8 \\
\hline
 3 & 0.1 & 0.5 & 3.46 & 0.81 &  ion  & $1.5\times10^7$ & $4.5\times10^{-5}$ &  3.6 & 8 \\
 {\small($m_\ion=25m_\lec$)} & & & & &  elec.& \idem   & \idem              &  89  & no power-law \\
\hline
 3 & 0.1 & 1   & 3.46 & 0.66 &  ion  & $1.5\times10^7$ & $2.8\times10^{-5}$ &  3.6 & 8 \\
 {\small($m_\ion=25m_\lec$)} & & & & &  elec.& \idem   & \idem              &  89  & 1.5 \\
\hline
 6 & 0.1 & 0   & 13.5 & 0.97 &  ion  & $8\times10^8$   & $7.5\times10^{-4}$ & 14   & 3.6 \\
 {\small($m_\ion=25m_\lec$)} & & & & &  elec.& \idem   & \idem              & 260  & 1.2 \\
\hline
 3 & 0.1 & 0   & 41.4 & 0.988&  ion  & $2\times10^8$   & $7.5\times10^{-4}$ &  83  & 1.5 \\
{\small(pairs)}& &     &      &      &  elec.& \idem   & \idem              & \idem& \idem \\
\hline
 3 & 0.1 & 0   & 6.9  & 0.93 &  ion  & $2\times10^8$   & $7.5\times10^{-4}$ & 7.5  & 3.6 \\
{\small($m_\ion=12m_\lec$)}& & & &   &  elec.& \idem   & \idem              &  83  & 1.5 \\
\hline
 6 & 0.1 & 0   & 6.9  & 0.93 &  ion  & $8\times10^8$   & $7.5\times10^{-4}$ & 7.1  &  5 \\
{\small($m_\ion=50m_\lec$)}& & & &   &  elec.& \idem   & \idem              & 260  &  1.5 \\
\end{tabular}
\end{table*}

\section{Introduction}\label{sec:intro}

Magnetic reconnection is a prime mechanism invoked to produce high-energy particles, 
radiation and high-energy flares, to launch large scale outflows, or to efficiently heat plasmas, 
in a variety of astrophysical objects.
It is a candidate to explain 
particle acceleration at pulsar wind termination shocks \citep{Kirk2003,Petri2007b,Sironi2011b},
the flat radio spectra from galactic nuclei and AGNs \citep{Birk2001} and from extragalactic jets \citep{Romanova1992},
GeV-TeV flares from the Crab nebulae \citep{Cerutti2012,Cerutti2012b,Cerutti2013},
flares in active galactic nuclei (AGN) jets \citep{Giannios2009}
or in gamma-ray bursts \citep{Lyutikov2006c,Lazar2009},
the heating of AGN and microquasar coronae and the observed flares \citep{Matteo1998,Merloni2001,Goodman2008,Reis2013,Romero2014,Zdziarski2014},
the heating of the lobes of giant radio galaxies \citep{Kronberg2004},
transient outflow production in microquasars and quasars \citep{deGouveia2005,deGouveia2010,Kowal2011,McKinney2012,Dexter2013},
gamma-ray burst outflows and non-thermal emissions \citep{Drenkhahn2002,McKinney2012b},
X-ray flashes \citep{Drenkhahn2002},
or soft gamma-ray repeaters \citep{Lyutikov2006b}.
In all these cases, it is crucial to know the amount of magnetic energy transferred to the particles during a reconnection event, 
the relative fraction distributed to ions and electrons, 
as well as the distribution in momentum space of the accelerated particles.
The aim of this manuscript is to shed light on these questions.
In the literature, several acceleration mechanisms by magnetic reconnection have been identified, 
that we briefly review in the remaining of this introduction.
Their relative importance depends on the plasma parameters and on the magnetic field geometry.

One acceleration mechanism occurs when particles are trapped in contracting magnetic islands, 
and thus accelerated by the induced electric field when they are reflected on the two approaching sides.
It can be efficient in collisionless plasmas \citep{Drake2006,Drake2010,Bessho2012}
or in collisional plasmas \citep{Kowal2011} where reconnection is fast because of turbulence\note{\citet{Drake2006}
for electrons in the solar corona; 
\citet{Drake2010} for ions at the heliospheric termination shock; 
\citet{Kowal2011} for e-/e+ in microquasars.}.
In non-relativistic plasmas, particle Larmor radii are smaller than the magnetic gradient scales,
particle motions are adiabatic inside and around the islands, and particle-in-cell simulations and analytical estimations agree that 
this mechanism produces power-law spectra,
with indexes $p=1.3$ or softer depending on the plasma $\beta$ and island aspect ratio \citep{Drake2006}.
In relativistic plasmas, the Larmor radii are of the order of the island scales, 
so that another analytical approach is to be employed
\citep{Bessho2012}, 
and there is no analytical expression for the resulting spectra.
PIC simulations in relativistic pair plasmas show that this mechanism
significantly contributes to the building of the high-energy population
\citep{Bessho2012,Sironi2014}, a result that we confirm to hold also for relativistic ion-electron reconnection.
\note{It saturates if the fire-hose instability threshold is reached.}

Another acceleration mechanism, also relying on the first order Fermi process and on stochasticity,
is bouncing motion of particles between the two inflows converging from both sides of the current sheet.
Energy is gained when the particle turns around, and is transferred by the motional electric field $\b{E} = -\b{v}\wedge\b{B}$ present in the 
inflow.
\citet{Drury2012} derives the power-law spectrum for non-relativistic particles:
$\dif n(v)/\dif v\propto v^{-p}$ with $v$ the velocity, $p=(r+2)/(r+1)$, where $r=n_\out/n_\ins$ is the compression ratio
that is not restricted to small values as in the case of shocks.
\citet{Giannios2010} derives the maximal Lorentz factor produced in the relativistic case. 
This mechanism does not rely on a direct acceleration by the reconnection electric field $E_\mathrm{rec}$ 
when particles are demagnetized
at the center of the diffusion region, but makes use of the motional electric field in the inflow.
It is thus efficient in non-relativistic and/or collisional plasmas \citep{Kowal2011} where direct 
acceleration by $E_\mathrm{rec}$ is known to be negligible.
It requires particles crossing the current sheet and bouncing on the other side, i.e., having a Larmor radius in 
the asymptotic field larger than the sheet width,
which is generally true only for pre-accelerated particles or hot inflows.
We show here that for cold inflows and relativistic setups, electrons and ions do not cross the current sheet,
and so do not undergo this acceleration mechanism.

A third acceleration mechanism is by the reconnection electric field $E_\mathrm{rec}$, 
which is initially induced by magnetic field flux variations, and sustained in steady or quasi-steady state 
by the non-ideal response of the plasma.
In the diffusion region, the condition $E>B$ for antiparallel reconnection, or $\b{E}\cdot\b{B}\neq0$ if there is a guide field, 
defines an acceleration region where particles can be freely accelerated and directly gain energy.
In any case, the reconnection electric field is alone responsible for the transfer of energy between the magnetic field and 
the particles, and thus obviously accelerate particles. But to which extent this kinetic energy is distributed 
between the bulk flow velocity of the outflows, their thermal energy, and a possible high-energy tail, 
as well as the properties of the high-energy tail, are open questions.
This mechanism is inefficient for non-relativistic reconnection because 
the acceleration zone has a too small length (along $z$ here) 
\citep{Drake2010,Kowal2011,Drury2012} and affects too few particles, but efficient under relativistic conditions 
where the larger reconnection electric field creates a wider acceleration zone \citep{Zenitani2001,Zenitani2007}.
It has indeed been found, with PIC simulations of relativistic reconnection, 
that power-law tails are produced through particle acceleration by $E_\mathrm{rec}$.
Several indexes are found, for example, measuring the index $p$ as $\dif n_\lec/\dif\gamma \propto \gamma^{-p}$
and retaining only relativistic particle-in-cell simulations that all concern pair plasmas:
\citet{Zenitani2001} (2D): $p=1$ for particles around the X-point and for the total spectra; 
\citet{Zenitani2007} (2D): $p=3.2$ and 2.4 at late times; 
\citet{Jaroschek2008} (2D, two colliding current sheets): power law; %
\citet{Sironi2011b} (2D, stripped pulsar wind): $p=1.4$ after the shock;
\citet{Cerutti2013} (2D): $p=3.8$;
\citet{Sironi2014} (2D without guide field): $p=4,\,3,\,2,\,1.5$ for inflow magnetizations $\sigma=1,\,3,\,10,\,30,\,50$ 
and a saturation above 50, and $p=2.3$ in 3D with $\sigma=10$.
On the other hand, \citet{Kagan2012} (3D) find a high-energy tail but interpret it as not having a power-law shape.
On the analytical side, \citet{Zenitani2001} present a toy model predicting power-laws, 
and \citet{Bessho2012} derive the spectrum of particles escaping from an antiparallel X-point (see Sect.~\ref{sec:other_acc_mechanism}).
Also, it is noticeable that the ultrarelativistic test particle simulations of \citet{Cerutti2012} produce very hard power-laws 
($p\sim -0.5$), with electrons accelerated along Speiser orbits without any stochasticity.
This diversity of results calls for a unified analysis of simulations with various initial configurations, that we aim to provide here.

Other acceleration mechanisms exist, especially far from the diffusion region.
A first example is stochastic acceleration in the turbulence associated with reconnection \citep{Kowal2011}. 
A second, important example, is at the magnetic separatrices that separates the non-reconnected/reconnected regions,
where plasma flows through a non-linear wave structure (see also Sect.~\ref{sec:astro_outlook}).
Particle acceleration should also occur at the dipolarization front.
Our simulation setup, with no localized initial perturbation, precludes the existence of these other mechanisms,
and instead we focus on acceleration close to the diffusion region and inside islands, which is likely to be 
important in relativistic setups.

This manuscript is dedicated to relativistic ion-electron plasmas, for which no study yet exists.
Such plasmas are likely present in AGN and microquasar coronae, 
in microquasar jets \citep{Kotani1994,Trigo2013}, or possibly in GRB and AGN jets.
Physical parameters will be discussed in Sect.~\ref{sec:astro_outlook}.



\note{Collisionless shocks, on their side, are also argued to produce high-energy particles in some of these systems, for example 
in internal or external shocks in gamma-ray burst jets \modif{(ref)}, in microquasar and AGN jets, 
or at pulsar wind termination shocks. 

The way by which collisionless shocks produce high-energy particles is now relatively clear. 
Simple estimations show that diffusive shock acceleration produces spectra 
that only depend on the shock compression ratio $r$ with 
$\dif n(\gamma)/\dif\gamma\propto\gamma^{-p}$, $p=(r+2)/(r-1)$ 
\citep[][for particles with $v\sim c$ and subluminal shocks]{Bell1978},
which  leads to $p=2.5$ for relativistic 2D shocks ($r=3$) or to $p=2$
for 3D shocks ($r=4$).
This is indeed confirmed by particle-in-cell simulations for ions and electrons
\citep{Spitkovsky2008a,Sironi2009,Sironi2011}, even if it is shown that
shock drift acceleration can produce harder power-laws ($p\sim2$ in 2D) or that 
the power-law is steeper near the critical angle, and disappears for superluminal shocks.
The power-law nature of the spectra is granted by the proportionality between 
the probability of escape at each particle crossing of the shock and 
the energy gain for each cycle.
\note{In details:
\citet{Achterberg2001} makes analytical and test particles estimations for ultra-relativistic unmagnetized shocks, and reach the conclusion that $p\sim2.4$.
\citet{Spitkovsky2008a} is the Fermi-at-last article, and shows that for relativistic unmagnetized pair shocks, we have $p\sim2.4$.
\citet{Sironi2009} is for relativistic, magnetized, pair shocks, and shows that DSA is the acceleration mechanism for subluminal shocks, but 
SDA becomes more important near the critical angle. The power law index is $p$ between 2.8 (DSA, 0 obliquity) and 2.3 (near the critical angle).
There is no high-energy tail for superluminal shocks.
\citet{Sironi2011} is for relativistic, magnetized, ion-electron shocks. It reaches the same conclusions concerning DSA and SDA as in pair shocks.
Weakly magnetized shocks ($\sigma<10^{-3}$) produce power-laws with $p\sim2.5$ for both electrons and ions
When the magnetization increases, so does the index: up to 5 for electrons and 3.5 for ions at $\sigma=1$ and $\theta=15$.
When the angle increases, the power-lax index dwindles (because SDA becomes important and is more efficient): from 5 to 2 for electrons 
and from 3.5 to 2 for ions (at fixed $\sigma=0.1$).
There is no high-energy tail for superluminal shocks.
On the theoretical side, for non-relativistic particles one has $dn(p)/dp\propto p^{-(r+2)/(r+1)}$, 
with $r=\rho_2/\rho_1 >1$ the compression ratio
(for strong NR shocks, $r=4$) and $p$ the momentum \citep{Drury1983}. 
For relativistic particles ($v\sim c$ in the equations), one has $dn(\gamma)/d\gamma\propto \gamma^{-(r+2)/(r+1)}$ 
with $\gamma$ the particle Lorentz factor,
for any (subluminal) magnetic field obliquity \citep{Bell1978}.
For relativistic shocks, $r=3$ and it gives $p=2.5$.
Note that $r$ seems to depend on the dimension, and $r=3$ at 2D, $r=4$ at 3D.}}

\section{Simulation setups}\label{sec:sim_setup}

\begin{table}[tbp]
\caption{\label{tab:param_physical_tearing}Parameters of the current sheet.
To each row in the table can correspond different background plasma parameters, and hence different simulations. 
The full simulation list is presented in Table~\ref{tab:param_magnetization}.
Here, the electron and ion temperatures are the same, denoted by 
$\Theta_\lec = T_\lec/(m_\lec c^2)$ and $\Theta_\ion = T_\ion/(m_\ion c^2)$. 
The electrons and ions counterstream with opposite velocities $\pm \beta_\lec c$ and associated Lorentz factors $\Gamma_\lec$.
The sheet half-width in units of ion inertial lengths is $L/d_\ion$, while in units of electron thermal Larmor radii it is $L/r_\mathrm{ce}$.
}
\centering
\begin{tabular}{c|c|c||c|c|c|c}
 $\frac{m_\ion}{m_\lec}$ & $\wce/\wpe$ & $L/d_\ion$ & $\Gamma_\lec \beta_\lec$ & $\Theta_\lec$ & $\Theta_\ion$ & $L/r_\mathrm{ce}$ \\
\hline
  1    & 3 & 2.5 & 0.53 & 2.40 & 2.40               & 1.6 \\
 12    & 3 & 0.5 & 0.53 & 2.40 & 0.2                & 1.1 \\
 25    & 1 & 0.5 & 0.20 & 0.25 & $1.0\times10^{-2}$ & 3.8 \\
 25    & 3 & 0.5 & 0.53 & 2.40 & $9.6\times10^{-2}$ & 1.6 \\
 25    & 6 & 1   & 0.70 & 10   & 0.4                & 1.5 \\
 50    & 6 & 0.7 & 0.60 & 10   & 0.2                & 1.5 \\
\end{tabular}
\end{table}

\subsection{The simulations}
\label{sec:the_simulations}

We perform 2D PIC simulations of magnetic reconnection,
mainly in an ion-electron plasma of mass ratio $m_\ion/m_\lec = 25$.
We also present one simulation for each value $m_\ion/m_\lec = 1$, 12, and 50.
We use the explicit PIC code \verb<Apar-T<, presented in \citet{Melzani2013}.
The simulations are the same as those of \citet{Melzani2014a}, to which we refer for details.
The initial state is a Harris equilibrium with a reversing magnetic field 
\begin{equation}
  \b{B}_\mathrm{rec} = \hat{\b{z}}\, B_0 \tanh\left( {x}/{L} \right),
\end{equation}
plus in some cases a guide field $\b{B}_\mathrm{G} = B_\mathrm{G} \hat{\b{y}}$.
The magnetic field $\b{B}_\mathrm{rec}$ is produced by a current sheet (abbreviated cs),
formed by counter-streaming ions and electrons following the density 
profile $n_\mathrm{cs}(x) = n_\mathrm{cs}(0)/\mathrm{cosh}^2 (x/L)$,
with bulk velocities $U_\lec$ and $U_\ion=-U_\lec$ in the $\pm y$ directions.
We denote the associated Lorentz factors by $\Gamma_\lec$ and $\Gamma_\ion$.
Each species follows a Maxwell-J\"uttner distribution.
The parameters of the current sheet are given in Table~\ref{tab:param_physical_tearing}.
They actually differ by $\sim 5\%$ from the values of the actual kinetic equilibrium, in order to speed up the otherwise slow initial phase.
We note that this initial perturbation is not localized in space, so that islands and X-point form everywhere along the current sheet. 

In addition, there is a background plasma at rest, with number density $n_\bg$ equal for ions and electrons,
and temperatures $T_{\bg,\ion}$ and $T_{\bg,\lec}$.

The free parameters are the characteristics of the background plasma ($n_\bg/n_\mathrm{cs}(0)$, $T_{\bg,\ion}$ and $T_{\bg,\lec}$);
the strength of the guide field $B_\mathrm{G}/B_0$;
the width of the magnetic field reversal in electron inertial lengths $L/d_\lec$; 
and the magnetization of the current sheet plasma with respect to the asymptotic magnetic field,
here expressed via $\wce/\wpe$ ($\wce$ is the electron cyclotron pulsation in the asymptotic magnetic field $B_0$, 
$\wpe$ is the electron plasma pulsation at the current sheet center at $t=0$). 
The background plasma magnetization results from the above variables.
The simulations, and the background plasma parameters and magnetizations, are listed in Table~\ref{tab:param_magnetization}.

We now give the definition of the magnetizations.
The inflow or background plasma magnetization for species $s$ is defined as the ratio of the energy flux in 
the reconnecting magnetic field to that in the inflowing particles \citep{Melzani2014a}:
\begin{equation}\label{equ:acc_sigma_s}
\begin{aligned}
 \sigma^\mathrm{hot}_s(B_\mathrm{rec}) &= \frac{E\times B_\mathrm{rec}/\mu_0}{n_{\mathrm{lab},s}\langle v\gamma m_sc^2\rangle_s} = \frac{B_\mathrm{rec}^2/\mu_0}{n_{\mathrm{lab},s}m_sc^2\Gamma_s h_{0,s}} \\
    &= \frac{\sigma^\mathrm{cold}_s(B_\mathrm{rec})}{\Gamma_s h_{0,s}}.
\end{aligned}
\end{equation}
Here we used $E = v_{\ins,s} B$ in the ideal inflowing plasma, and the relation 
$\langle v\gamma\rangle_s = h_{0,s}\Gamma_s v_{\ins,s}$ \citep{Melzani2013}, where $\langle\cdot\rangle_s$ 
denotes an average over the momentum distribution function,
with $h_{0,s}$ the comobile enthalpy, and $\Gamma_s=(1-v_{\ins,s}^2/c^2)^{-1/2}$.
Also, $n_{\mathrm{lab},s}$ is the lab-frame particle number density,
and $\sigma^\mathrm{cold}_s$ the magnetization of the plasma without taking into account temperature effects or relativistic 
bulk motion:
\begin{equation}\label{equ:acc_sigma_s_cold}
 \sigma^\mathrm{cold}_s(B) = \frac{B^2}{\mu_0n_{\mathrm{lab},s}m_sc^2}.
\end{equation}
Finally, the total magnetization of the plasma is
\begin{equation}\label{equ:sigma_tot}
 \sigma_{\ion+\lec}(B_\mathrm{rec}) = \frac{B_\mathrm{rec}^2/\mu_0}{\sum_sn_{\mathrm{lab},s}m_sc^2\Gamma_s h_{0,s}}
    = \frac{\sigma^\mathrm{cold}_\ion(B_\mathrm{rec})}{\sum_s\Gamma_s (m_s/m_\ion) h_{0,s}}.
\end{equation}
For non-relativistic temperatures ($h_{0,s}\sim 1$) and non-relativistic inflow velocities ($\Gamma_s\sim 1$),
the total magnetization reduces to $\sigma_{\ion+\lec} \simeq \sigma^\mathrm{cold}_\ion / (1+m_\lec/m_\ion) \simeq \sigma^\mathrm{cold}_\lec / (1+m_\ion/m_\lec)$.

The relativistic Alfv\'en velocity in the inflow is defined by 
\begin{equation}\label{equ:def_Alfven_speed}
 \left(\frac{V^\mathrm{R}_\mathrm{A,in}}{c}\right)^2 = \frac{\sigma_{\ion+\lec}(B_\tot)}{1+\sigma_{\ion+\lec}(B_\tot)},
\end{equation}
with $B_\tot = (B_0^2+B_\mathrm{G}^2)^{1/2}$, and $\sigma_{\ion+\lec}(B_\tot)$ the comobile plasma total magnetization
(Eq.~\ref{equ:sigma_tot} with $\Gamma_s=1$ and $n_{\mathrm{lab},s}$ replaced by the comobile number density).
When there is a guide field, it is relevant to project the Alfv\'en velocity~\ref{equ:def_Alfven_speed} 
in the plane of the reconnecting magnetic field: $V^\mathrm{R}_\mathrm{A,in}\cos\theta$ with $\tan\theta = B_\mathrm{G}/B_0$.
This is needed to correctly normalize the reconnection electric field \citep{Melzani2014a}.


\begin{figure*}[tb]
 \centering
 \def\svgwidth{\textwidth}
 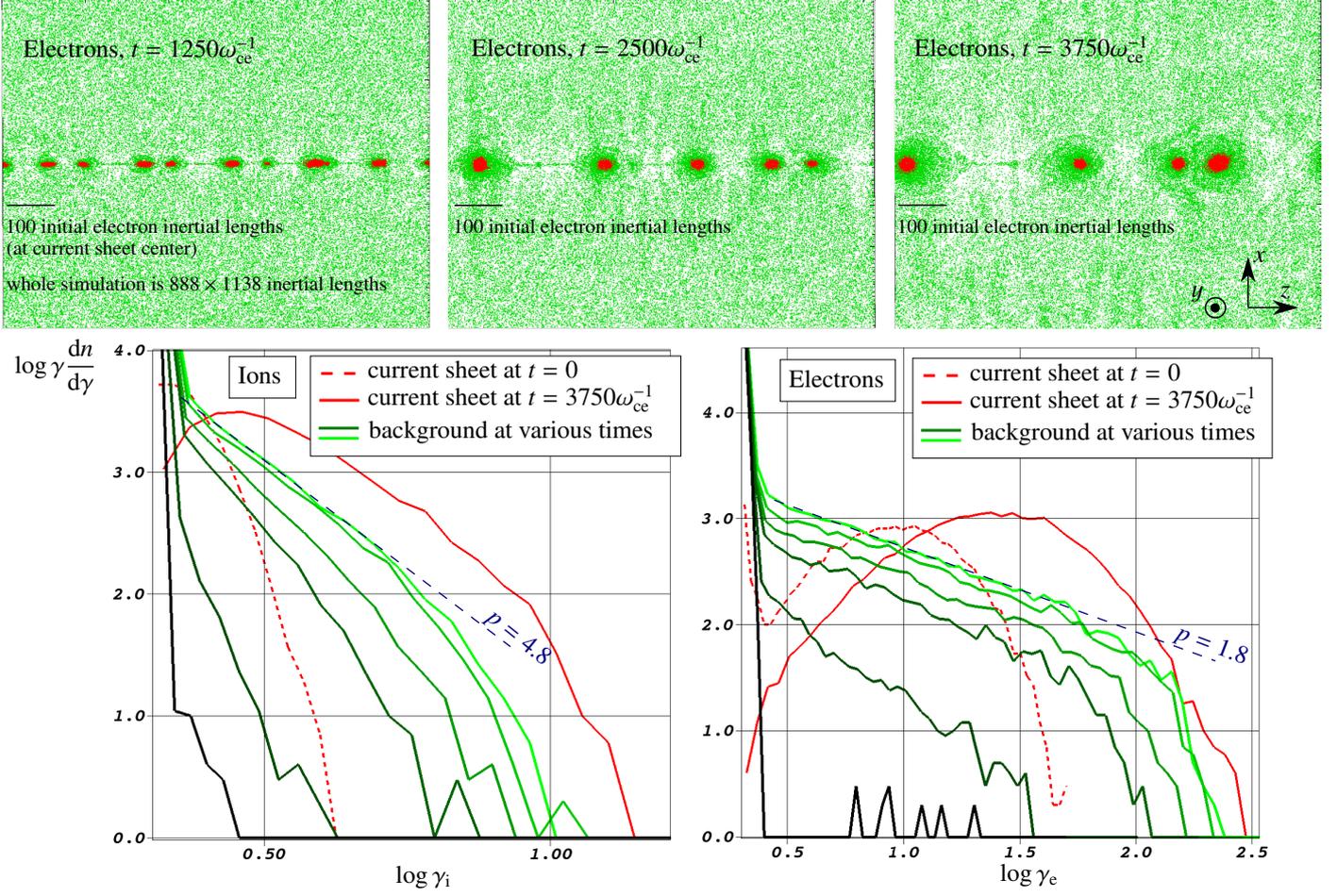
 \caption{\label{fig:histo_electrons_large_simulation}
 Data from the main simulation (Sect.~\ref{sec:main_case}), with a background magnetization respectively for ions and electrons
 $\sigma^\mathrm{hot}_{\ion,\,\lec}=3.6,\,83$.
 Top: snapshots of a random selection of electrons in the whole simulation domain. Red particles are inside the current sheet at $t=0$, 
 green ones are outside.
 Bottom: Lorentz factor distributions. Red (green) curves concern the red (green) population.
 For the green curves, times are ordered as dark to light green, with values 0, 750, 1500, 2250, 3000, 3750$\wce^{-1}$, i.e., one curve every 
 $750\wce^{-1} = 250\wpe^{-1} = 50\wpi^{-1} = 30\wci^{-1}$. The blue dashed line indicates the final power-law slope 
 of the background accelerated particles.
 }
\end{figure*}

\begin{figure}[tb]
 \centering
 \def\svgwidth{\columnwidth}
 \begin{tiny}
 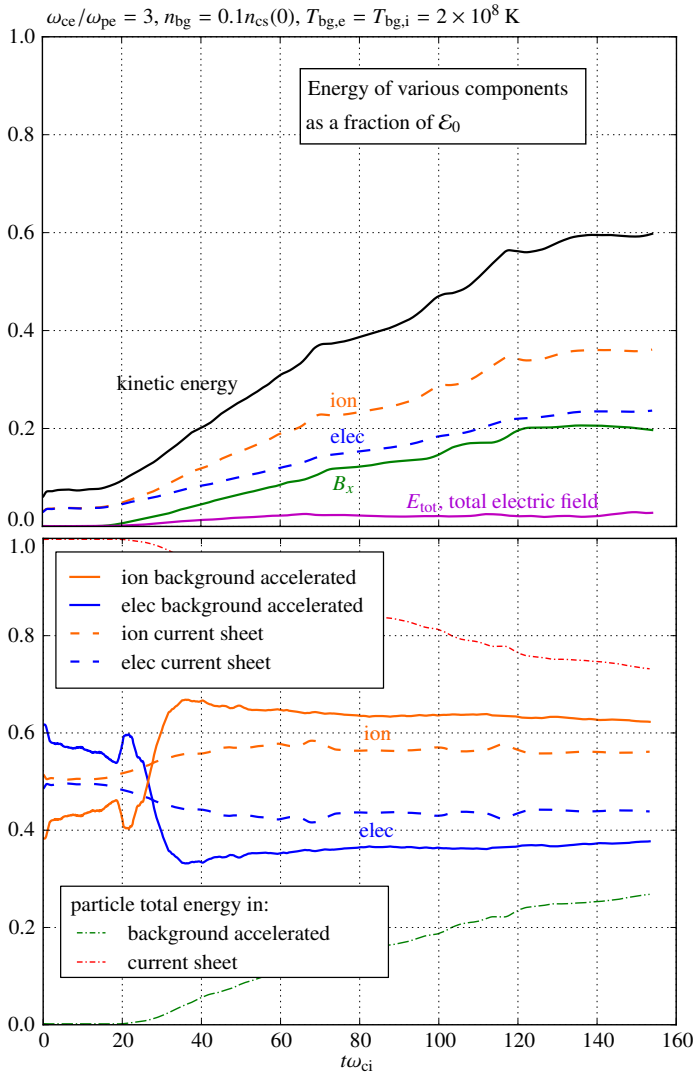
 \end{tiny}
 \caption{\label{fig:energy}
 Energy distribution for the main simulation (Sect.~\ref{sec:main_case}), 
 with a background magnetization respectively for ions and electrons
 $\sigma^\mathrm{hot}_{\ion,\,\lec}=3.6,\,83$.
 Top: Energy in the total electric field $E_\tot$, in the $x$ component of the magnetic field,
 and in the particles (also decomposed into the ion and electron contributions).
 These energies are computed on a fixed area, that corresponds to the location of the background particles, or field lines,
 that reach the current sheet before the end of the simulation. The total initial energy in this area is $\mathcal{E}_0$.
 We note that the energy in $E_\tot$ is mostly that in $E_y$.
 Also, the energy of $B_y$ is 0.5\% of that of $B_x$.
 Bottom: Orange and blue curves are the energy repartition between ions and electrons for the current sheet particles
 (dashed), and for the background particles that have been accelerated (solid). They are normalized so that their sum is 1.
 The red and green curve show the energy repartition between current sheet particles and background accelerated particles.
 }
\end{figure} 

\subsection{Resolution and domain size}
\label{sec:resolution_tests}

The numerical resolution is set by the number of cells $n_x$ per electron inertial length $d_\lec$,
by the number of timesteps $n_t$ per electron plasma period $2\pi/\wpe$, 
and by the number of computer particles (the so-called superparticles) per cell $\rho_{\mathrm{sp}}$.
The quantities $d_\lec$, $\omega_\pe$, and $\rho_{\mathrm{sp}}$ are defined at $t=0$ at the center of 
the current sheet.
Here we take $n_x=9$ and $n_t=150$ (except for $\wpe/\wce=6$ where $n_t=250$).
We checked by doubling $n_x$ and $n_t$ that the particle distributions, the energy repartition, 
particle mixing, or the maximal Lorentz factors,
are not affected by the resolution.

Concerning the number of superparticles per cell at the center of the current sheet, 
we use $\rho_{\mathrm{sp}}=1090$ for $n_\bg/n_\mathrm{cs}(0)=0.3$,
and $\rho_{\mathrm{sp}}=1820$ for $n_\bg/n_\mathrm{cs}(0)=0.1$, except for $m_\ion/m_\lec=50$
where $\rho_{\mathrm{sp}}=910$.
This corresponds, for the case $n_\bg/n_\mathrm{cs}(0)=0.1$, to 1650 electron and ion superparticles per cell 
for the plasma of the current sheet, 
and to 170 for the background plasma. 
We stress in \citet{Melzani2013,Melzani2013b} that because of their low numbers of superparticles 
per cell when compared to real plasmas, PIC simulations present high levels of correlations and collisionality,
and thus thermalize faster. 
In the same line of thoughts, \citet{Kato2013} and \citet{May2014} show that because of these enhanced correlations, 
high-energy particles are slowed down quickly in PIC plasmas.
One should thus ensure that collisionless kinetic processes remain faster than collisional effects,
essentially by taking a large enough number $\Lambda^\mathrm{PIC}$ of superparticles per Debye sphere
and per inertial length sphere, the former constraint being more restrictive.
For example, with $\Theta_\lec=2.4=1.4\times10^{10}\,\mathrm{K}/(m_\lec c^2)$
the electron Debye length is 20 cells large, and we have initially at the center of the current sheet: 
$\Lambda^\mathrm{PIC} \sim 1820\times20\times20 = 7.3\times10^5$ superparticles.
For a background plasma with $T_\bg=2\times10^8$\,K, we have $\Lambda^\mathrm{PIC}=133$.
We performed a simulation with twice less superparticles per cell, 
and saw no difference, especially concerning particle distributions, energy repartition, 
particle mixing, or maximal Lorentz factor.
It indicates that we are not affected by~$\rho_\mathrm{sp}$.

Boundaries are periodic along $z$ and $y$, reflective along $x$.
The simulation with $m_\ion/m_\lec=50$ uses a domain size of $8000\times10240$ cells.
The number of cells of the simulations with other mass ratios is $4100\times6144$, 
corresponding to $455\times683$ initial electron inertial lengths $d_\lec$,
with typically $4\times10^9$ superparticles.
We performed a simulation with a twice smaller domain along $z$: 
particle distributions are identical as long as there is a significant number of islands and X-points
in the domain ($\geq 4$), but differ afterward. In the smaller simulation, the distribution cutoff is at lower energies, 
and the power-laws are steeper (softer).
We also performed a simulation with a larger domain 
($8000\times10240$ cells, i.e., $888\times1138$ inertial lengths $d_\lec$)
and otherwise identical parameters:
the electron distribution saturates identically to the $4100\times6144$ case, but the ion distribution
reaches a harder final state. It indicates that our domain size and simulation duration are large enough for electrons, 
but possibly not for ions. The latter may build harder spectra and reach higher energies in real systems.

\subsection{Diagnostics}

We initially select, uniformly in space, 
of the order of 200\,000 particles (out of the 4 to 14 billions in total), and write their positions, 
velocities, as well as the magnetic and electric fields they undergo, once every a few timesteps.
The visualization of these data is performed with the visualization software VisIt \citep{HPV:VisIt}.
We divide the followed particles into two populations: those that are initially in the current sheet
(colored in red), and those initially outside (marked in green). Said otherwise, red particles are those 
satisfying 
\begin{equation}
 \text{distance from middle plane at }t=0 < 2L,
\end{equation}
while green particles satisfy 
\begin{equation}
 \text{distance from middle plane at }t=0 > 2L.
\end{equation}
Changing the limiting length from $2L$ to between $1.5L$ and $3L$ has been checked not to influence the presented results.
An example is shown in Fig.~\ref{fig:histo_electrons_large_simulation}.
As we will show, these two populations almost do not mix spatially, and undergo very different 
acceleration mechanisms, resulting in completely different particle energy distributions.
Particles from the background plasma, accelerated by the reconnection, are expected to dominate in number and energy 
for very large systems. This is why we focus more on the green population.

\begin{figure*}[tb]
 \centering
 \def\svgwidth{\textwidth}
 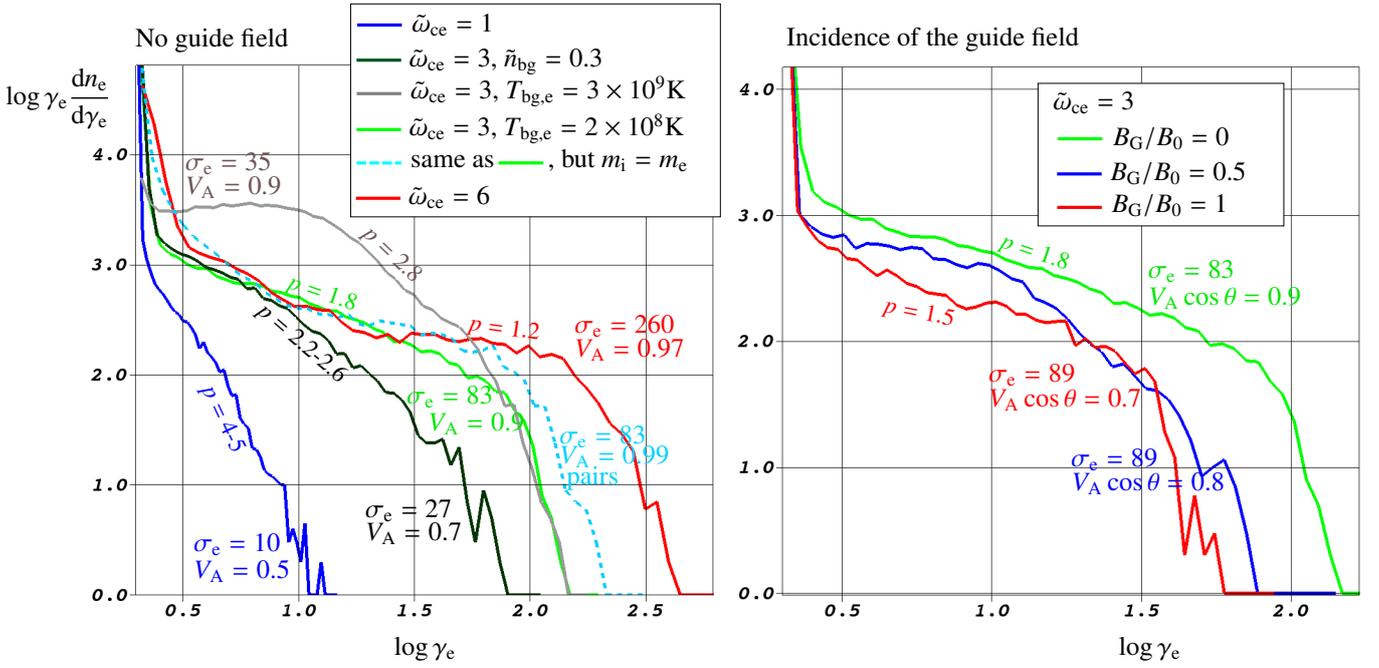
 \caption{\label{fig:histo_lec_summary_all}Lorentz factor distributions for background electrons (green population), 
for various simulations with $m_\ion/m_\lec=25$ or 1, in the final state.
 Times are $t = 600\wpe^{-1} = 1800\wce^{-1}$ for all simulations, except for $\wce/\wpe=1$ ($t = 1350\wpe^{-1} = 1350\wce^{-1}$),
 and for $\wce/\wpe=6$ ($t = 600\wpe^{-1} = 3600\wce^{-1}$).
 Notations are abbreviated: $\tilde{\omega}_\ce=\wce/\wpe$, $\tilde{n}_\bg=n_\bg/n_\cs(0)$,
 $\sigma_\lec$ stands for $\sigma_\lec^\mathrm{hot}(B_\mathrm{rec})$,
 and $V_\mathrm{A}$ for $V^\mathrm{R}_\mathrm{A,in}\cos\theta/c$ (with $\theta=\arctan B_\mathrm{G}/B_0$). 
 Unless specified, $\tilde{n}_\bg=0.1$. 
 The power-law indexes are $p$.
 The precise setups are reported in Table~\ref{tab:param_magnetization} for each $\sigma_\lec$.
 The light-green curve ($\sigma_\lec = 83$, $V_\mathrm{A}=0.9$) is the same on the left and right panel, 
 and is the final state of the simulation of Fig.~\ref{fig:histo_electrons_large_simulation}.
 }
\end{figure*}

\section{Results}\label{sec:results}

We first present results from a reference simulation in Sect.~\ref{sec:main_case},
and then study modifications due to varying the background particle number density, magnetic field, or temperature
without considering a guide field in Sect.~\ref{sec:case_study_Bg0}.
The consequences of a guide field are explored in Sect.~\ref{sec:case_study_Bg}.

\subsection{Main case}\label{sec:main_case}

We start by presenting the results of the simulation with $\wce/\wpe=3$, $n_\bg=0.1n_\mathrm{cs}(0)$, 
$T_{\bg,\ion} = T_{\bg,\lec} = 2\times10^8$\,K, resulting in a magnetization in the background plasma 
$\sigma^\mathrm{hot}_{\ion,\,\lec}=3.6,\,83$ for ions and electrons, respectively
(defined by Eq.~\ref{equ:acc_sigma_s}, see also Tables~\ref{tab:param_magnetization} and~\ref{tab:param_physical_tearing}),
and in an inflow Alfv\'en speed $V^\mathrm{R}_\mathrm{A,in}=0.88c$.

The background electron distribution (the green curves in Fig.~\ref{fig:histo_electrons_large_simulation})
starts rising and taking a power-law shape around $t=500\wce^{-1}$, which corresponds to the 
starting of the reconnection instability. 
What happens is that the reconnection electric field $E_\mathrm{rec}$ spreads in the background plasma and sets the particles into motion in an 
$E\times B$ drift directed toward the current sheet. More and more background particles thus pass in the current sheet,
where they are demagnetized and able to gain energy via $E_\mathrm{rec}$.
The power-law component thus comprises more and more particles. 
After gaining energy in the acceleration zone, particles accumulate around the magnetic islands and swirl around them. 
They are subsequently accelerated when islands merge and contract.
The power-law index passes from a soft initial value of $p = -\dif \log n(\gamma) / \dif \log\gamma = 3$
to a harder final value that converges to $p\sim 1.8$.
We stopped the simulation at $t=3750\wce^{-1}$, when there are still enough islands and X-points so that we are not affected by boundaries 
(Sect.~\ref{sec:resolution_tests}).

Concerning the current sheet electrons, their distribution is shown at $t=0$ by a red dashed line
in Fig.~\ref{fig:histo_electrons_large_simulation}, and consists then in a thermal hot plasma.
When the reconnection instability starts, this population is heated by the formation and contraction of islands.
This heating slowly progresses at later times as islands merge, to result in the solid red curve of Fig.~\ref{fig:histo_electrons_large_simulation}.

Concerning ions, their background magnetization is only slightly relativistic ($\sigma^\mathrm{hot}_{\ion}=3.6$).
The current sheet population is heated, while the background population distribution is power-law like, 
with a final index $p\sim 4.8$. This is similar to non-relativistic simulations where all species form steep spectra.

It is evident from the top panels of Fig.~\ref{fig:histo_electrons_large_simulation}
that the green electrons from the background plasma
do not penetrate deep inside the islands, 
and on the other hand that the red electrons initially from the current sheet do not escape from the islands, even when they merge.
The two populations thus remain almost separated.
This is because particles from the background plasma are scattered by the strong magnetic field structure 
surrounding the islands, and thus swirl around these field lines, performing circles around the islands 
but not reaching the inside.
On their side, red particles from the islands cannot escape because of the very same magnetic field structure.
This remains true for ions, but less so because of their larger Larmor radius.
Figure~\ref{fig:islands_mixing} illustrates this population separation for several simulations.

The energy repartition between fields and particles is shown in Fig.~\ref{fig:energy} (top).
This energy is computed over a fixed rectangle in space,
defined to include all particles that will reach the center of the current sheet before the end of the simulation.
It thus excludes regions that, because of the finite simulation length, are never in contact with the current sheet.
Energies are normalized by $\mathcal{E}_0$, the total initial energy in this area, which is to 
$\sim 90\%$ the energy in the magnetic field.
The energy in the magnetic field is transferred to the kinetic energy of the particles 
($0.6\mathcal{E}_0$ in the final state), to the reconnected magnetic field $B_x$ ($0.2\mathcal{E}_0$), 
and to the reconnection electric field $E_y$ ($0.03\mathcal{E}_0$).
The energy in $B_y$ and in $E_x$, $E_z$ is far smaller ($\sim0.005\mathcal{E}_0$).
A first conclusion is that the fraction of dissipated magnetic energy is large.
Table~\ref{tab:energy_distribution} presents this analysis for several simulations. 
It shows that the fraction of dissipated 
magnetic energy is even larger at larger inflow magnetization.
A second important aspect is the energy repartition between ions and electrons.
In Fig.~\ref{fig:energy} (bottom) we show this repartition for the background particles that have been accelerated,
i.e., for the particles of the tails in Fig.~\ref{fig:histo_electrons_large_simulation} (green curves).
The ions weight as 60\% of this kinetic energy, the electrons 40\%, and this ratio remains constant with time.
The same repartition roughly holds for particles in the current sheet (red population).
Table~\ref{tab:energy_distribution} shows that this repartition holds for various simulations
with different magnetizations and mass ratios, provided that there is no guide magnetic field.
We note that the percentages given in Table~\ref{tab:energy_distribution} are not sensitive to
the specific time when they are determined. We obtain essentially
the same percentages if we repeat the analysis but consider only
those particles (and their associated rectangular region and energy $\mathcal{E}_0$)
that reach the current sheet before half of the total simulation time.

\subsection{Case studies, no guide field}\label{sec:case_study_Bg0}

\subsubsection{Influence of the background plasma density}

We now compare the main case with 
a simulation with identical parameters except for the 
background plasma number density: 
$n_\bg = 0.3n_\cs(0)$ instead of 0.1, 
resulting in a smaller magnetization $\sigma^\mathrm{hot}_{\ion,\,\lec}=1.2,\,27$ in the background.
The corresponding inflow Alfv\'en speed is $V^\mathrm{R}_\mathrm{A,in}=0.73c$.
The evolution is similar: weak mixing of the two populations, heating of the particles from the current sheet, 
and acceleration of the particles from the background to form a power-law.
The power-law for electrons has a final index $p$ between 2.2 and 2.6 (Fig.~\ref{fig:histo_lec_summary_all}),
that for ions around $p\sim 8$.
This is softer than in the $n_\bg = 0.1n_\cs(0)$ case, 
which is expected because a higher background plasma density implies a lower magnetization, 
and as we show here \citep[see also][]{Sironi2014}, softer power-laws.
It is however noticeable that more magnetic energy is transferred to the particles: 
the kinetic energy is 74\% of the total energy, 
while it is only 62\% for the case with $n_\bg = 0.1n_\cs(0)$.


\subsubsection{Influence of the inflow magnetization}

We vary the asymptotic magnetic field strength by varying the parameter $\wce/\wpe$ of the Harris equilibrium.
Increasing this parameter results in a higher magnetic field, and in a hotter current sheet plasma in order to maintain the pressure balance. 
It allows to probe different background plasma magnetizations.
Our results indicate harder power-laws at higher magnetizations
(Fig.~\ref{fig:histo_lec_summary_all} for the simulations with $\wce/\wpe=1$, 6,   
to be compared also with the simulation $\wce/\wpe=3$, $T_{\bg,\lec}=2\times10^8$\,K).
The comparison is as follows:
\begin{itemize}
 \item The case at low inflow magnetization ($\wce/\wpe=1$, $\sigma^\mathrm{hot}_{\ion,\,\lec}=0.4,\,9.9$,
 $V^\mathrm{R}_\mathrm{A,in}=0.53c$) 
presents a power-law-like spectrum for background accelerated electrons (green population) 
with a large index, between 3 and 4, thus decreasing fast and reaching $\gamma_\mathrm{max} \sim 10$.
The ion distribution is not power-law like and is very steep.
The final kinetic energy is 48\% of the total initial energy.
 \item The intermediate case ($\wce/\wpe=3$, $\sigma^\mathrm{hot}_{\ion,\,\lec}=3.6,\,90$,
 $V^\mathrm{R}_\mathrm{A,in}=0.88c$) 
presents a power-law-like spectrum 
for the background particles with, for electrons $p\sim 1.5$-2 and $\gamma_\mathrm{max} \sim 100$,
for ions $p\sim 5.8$.
The final kinetic energy is 62\% of the total initial energy.
 \item The most magnetized and relativistic case ($\wce/\wpe=6$, $\sigma^\mathrm{hot}_{\ion,\,\lec}=14,\,260$,
 $V^\mathrm{R}_\mathrm{A,in}=0.97c$) 
exhibits a very flat spectrum for background accelerated electrons, 
with an index around 1.2, and reaches $\gamma_\mathrm{max} \sim 300$.
Background accelerated ions have a power-law distribution with index $p\sim3.6$, 
which is interestingly close to the index for electrons at the same magnetization (the simulation with 
$\wce/\wpe=1$, where $\sigma_\lec^\mathrm{hot}=9.9$ and $p\sim4.5$), 
and highlights the relevance of $\sigma_s$ of each species to characterize the power-law.
The final kinetic energy is 73\% of the total initial energy.
\end{itemize}
For both electrons and ions, this emphasizes the fact that 
only relativistic reconnection setups 
(i.e., $\sigma^\mathrm{hot}_s > 1$ for each species $s$) 
can produce power-laws, with harder indexes for higher magnetizations. 
%

\subsubsection{Influence of the inflow temperature}

Coming back to the main simulation with $\wce/\wpe=3$, $n_\bg=0.1n_\cs(0)$, $T_{\bg,\lec} = T_{\bg,\ion} = 2\times10^8$\,K,
we now increase the background temperature of the electrons to reach $T_{\bg,\lec} = 3\times10^9$\,K 
(giving $\sigma^\mathrm{hot}_{\ion,\,\lec}=3.6,\,35$, $V^\mathrm{R}_\mathrm{A,in}=0.88c$), 
which is almost the temperature of the current sheet electrons ($\Theta_\lec=2.4=1.4\times10^{10}\,\mathrm{K}/(m_\lec c^2)$).
Electrons from the background plasma already have a high energy when reaching the acceleration region,
and their initial energy is then a significant fraction of the energy gain furnished by $E_\mathrm{rec}$. 
As a consequence, the power-law is less visible (Fig~\ref{fig:histo_lec_summary_all}, gray curve).
However, as time goes by and as more and more particles from the hot background are accelerated, we expect it to 
dominate more and more the particle distribution.
Its index is $p=2.8$, softer than the colder case. 
This is expected because a relativistic temperature decreases the plasma magnetization
from 89 to 35 here, and we do have an index close to the one for $n_\bg=0.3n_\cs(0)$, which had a similar magnetization
($\sigma^\mathrm{hot}_\lec=27$, $p \sim 2.5$).

Background accelerated ions have a power-law distribution with index $p\sim 6.5$.
This is close to their index in 
the simulation with $\wce/\wpe=3$, $T_{\bg,\lec}=T_{\bg,\ion}=2\times10^8$\,K, which is identical except for the initial electron 
temperature ($p\sim5.8$). It shows that electrons have a weak influence on ions.

\begin{figure}[tb]
 \centering
 \def\svgwidth{\columnwidth}
 \begin{tiny}
 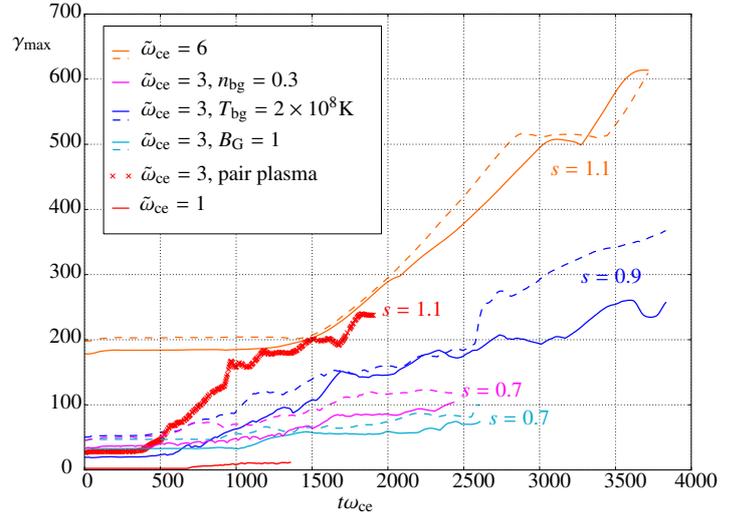
 \end{tiny}
 \caption{\label{fig:max_Lorentz_factor}Maximum Lorentz factor of the background particles for various simulations with $m_\ion/m_\lec=25$ or 1.
 Solid lines are for electrons, dashed lines are for ions and represent $m_\ion/m_\lec\times\gamma_\mathrm{i,max}$.
 Also, $\tilde{\omega}_\ce = \wce/\wpe$.
 A log-log plot shows that the Lorentz factors grow as $t^s$, with the index $s$ shown in the figure.
 }
\end{figure}

\subsection{Case studies, guide field}\label{sec:case_study_Bg}

We now analyze simulations with a guide magnetic field $\b{B}_\mathrm{G}=B_\mathrm{G}\hat{\b{y}}$.

Figure~\ref{fig:histo_lec_summary_all} (right) presents the final electron distributions
from simulations with $B_\mathrm{G}=0$, $0.5B_0$ and $B_0$. 
In the intermediate case ($B_\mathrm{G}=0.5B_0$), the spectrum of the background accelerated electrons shows no clear power-law, 
but extends over a range similar to the no guide field case.
In the strong guide field case ($B_\mathrm{G}=B_0$), the spectrum of the same electron population seems flatter, 
with a power-law index of $\sim 1.5$ (over a small range).
The background plasma magnetization in the three simulations is similar
(the guide field is not included in this magnetization parameter, and should not be, because it is not converted into particle energy).
The harder particle distribution should thus be explained by other means.
We recall that quite generally, the particles accelerated at the X-point are slowly deflected toward the 
reconnection exhausts by the $x$ component of $B$. 
But $B_x$ vanishes at the X-point and increases when going away from it, so that the further from the X-point a particle 
enters the diffusion region, the faster it will be deflected and the less it will be accelerated. 
The area where an efficient acceleration occurs is thus limited along $z$ by the increase of $B_x$.
But with a guide field, this efficient acceleration region is extended along $z$ \citep[as is shown in][]{Melzani2014a},
because accelerated particles are guided by the guide field and prevented from being deviated by $B_x$.
Background particles are more accelerated, and a flatter spectrum is indeed expected.

Because of their larger Larmor radii, background accelerated ions are less affected by the guide field.
They present a power-law distribution with index $p=8$ for both guide field strengths.

Table~\ref{tab:energy_distribution_guidefield} shows the energy repartition for the guide field simulations.
In both case, the final kinetic energy is $\sim44\%$ of the total initial energy 
(we do not include the guide field $B_y$ in this initial energy because it cannot be transferred to the particles,
and indeed varies by less than a few percent during the simulation).
This is less than in the $B_\mathrm{G}=0$ case, where this fraction is $62\%$.
The kinetic energy repartition between accelerated ions and electrons 
is 46\%/54\% (ions/electrons) for $B_\mathrm{G}=0.5B_0$, and 33\%/67\% for $B_\mathrm{G}=B_0$.
This contrasts with the 60\%/40\% ratio at $B_\mathrm{G}=0$, since here electrons get more energy than ions.


\section{Summary and discussion}\label{sec:ccl}

\subsection{Summary}

We study the production of high-energy particles by magnetic reconnection in relativistic ion-electron plasmas
based on the same 2D PIC simulation data presented by \citet{Melzani2014a}. 
The variety of parameters employed (particle density, temperature, or magnetic field 
in the background plasma, guide field, mass ratio) allows to grasp important aspects of this problem.
In all cases particles can be divided into two populations that only weakly mix:
(i) Particles initially inside the current sheet are trapped inside the magnetic islands as soon as they form during the tearing 
instability, and remain trapped by the strong circling magnetic structure, even after many island merging events. 
They are heated by the contraction of the islands.
(ii) Particles from the background plasma $E\times B$ drift toward the diffusion region, 
where either $E<B$ in the no guide field case, or $\b{E}\cdot\b{B}\neq0$ in the guide field case, 
allows them to be directly accelerated. As they escape along the reconnection exhaust, they cannot penetrate inside the island 
because of the strong magnetic structure surrounding them, and circle around at the periphery, 
where they can further gain energy. 

Particles of population (ii) tend to form a power-law whenever their magnetization is larger than unity 
and the inflow Alfv\'en speed is not too small
(Figs.~\ref{fig:histo_electrons_large_simulation}
and~\ref{fig:histo_lec_summary_all}), 
though sometimes not with a clear and unique slope.
The indexes depend on the temperature, particle density, and magnetic field in the background plasma, 
and on the guide field strength. 
With no guide field, results suggest that 
the power-law index for species $s$ depends mainly on the background plasma Alfv\'en speed $V_\mathrm{A,in}^\mathrm{R}$, and 
on the background plasma magnetization for species $s$, independent of whether it is due to 
the magnetic field strength, a lower particle density, or a relativistic temperature.
A higher magnetization leads to a harder power-law:
for the electrons, $p = -\dif\log n(\gamma)/\dif\log\gamma = 4.5,\,2.4,\,2.8,\,1.7,\,1.2$ respectively for magnetizations 
$\sigma^\mathrm{hot}_\lec = 10,\,27,\,35,\,89,\,260$ (see Table~\ref{tab:param_magnetization}).
This is expected for reasons exposed in Sect.~\ref{sec:discussion_cond_high_energy}.
These indexes are harder than for collisionless shock acceleration, where $p>2$ \citep{Bell1978,Sironi2011}.
Ions have a magnetization $m_\ion/m_\lec$ times smaller than electrons (for identical or non-relativistic temperatures).
As expected, they behave non-relativistically for low magnetizations $\sigma^\mathrm{hot}_\ion$: no power law at $\sigma^\mathrm{hot}_\ion=0.4$;
steep ones for $\sigma^\mathrm{hot}_\ion = 1.2,\,3.6$ ($p=8,\,5.5$);
and beginning of formation of significant power-laws at higher magnetization: 
$p=3.6$ for $\sigma^\mathrm{hot}_\ion = 14$, 
mimicking the values reached for electrons at the same $\sigma_\lec$.

The presence of a weak guide field deforms the power-law, and the presence of a strong guide field
allows for a slightly harder spectrum (Fig.~\ref{fig:histo_lec_summary_all}, right) 
because it allows particles to stay longer in the acceleration region by guiding them in the 
direction of the reconnection electric field.

\begin{figure}[tb]
 \centering
 \def\svgwidth{\columnwidth}
 \begin{tiny}
 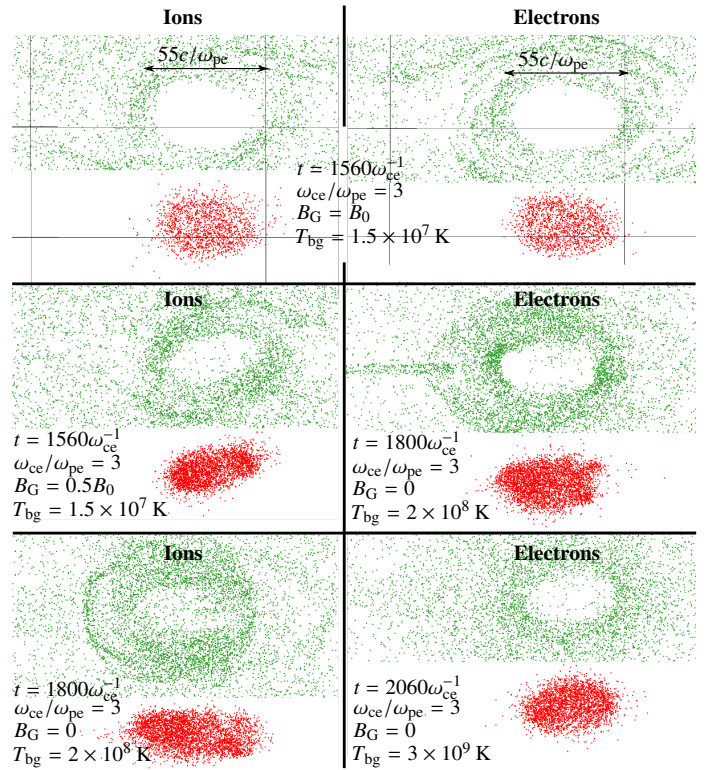
 \end{tiny}
 \caption{\label{fig:islands_mixing}Particle mixing in the islands. Each panel is a zoom around a magnetic island and shows, 
 in green, particles initially outside of the current sheet, and in red and shifted below for clarity, particles initially 
 located inside the current sheet. 
 Each snapshot is taken at the end of the simulations, and these islands result from the merging of many small and then larger islands
 (from around 20 islands at the end of the linear phase of the tearing instability, to 3 big islands at the end);
 yet the two particle populations remain separated. The mass ratio is 25 in all cases.
 }
\end{figure} 

The degree of mixing between the two populations (i) and (ii) essentially depends on the ratio of the Larmor radii 
of the particles in the magnetic field surrounding the islands, to the island radius. 
A hotter background temperature implies more mixing, and so does a weaker guide field.
Also, ions have larger Larmor radii and penetrate more easily inside the islands.
This is illustrated in Fig.~\ref{fig:islands_mixing}.
We stress that even for high background electron temperatures (bottom right panel), 
the two electron populations remain clearly separated.

The total particle distribution is the sum of populations (i) and (ii), and depends on their relative importance.
We underline that our simulations are limited in time by the box size. In reality, larger times can be reached
and more background particles can be accelerated, so that the background-accelerated population, and its power-law nature,
will dominate in the end. 
It calls for care when interpreting PIC particle distributions: 
either very long simulations (demanding also large domains) or the proposed decomposition should be used.

The fraction of magnetic energy converted into kinetic energy is larger at larger inflow magnetization: 
the final kinetic energy rises to 
48\%, 62\%, 73\% of the total initial energy, 
for respective inflow magnetizations $\sigma_\lec^\mathrm{hot} = 9.9$, 83, 260 (see Table~\ref{tab:energy_distribution})
at $m_\ion/m_\lec=25$.
This fraction is lower with a guide magnetic field ($\sim44\%$, Table~\ref{tab:energy_distribution_guidefield}).

The energy repartition between accelerated ions and electrons from the background plasma 
depends mainly on the strength of the guide magnetic field.
With no guide field, it is roughly 60\% for ions and 40\% for electrons,
with variations within 3\% when varying the background temperature, magnetization and Alfv\'en speed.
With a guide field of $0.5B_0$ and $B_0$, the ion/electron repartition becomes 
46\%/54\% and 33\%/67\%, respectively, with electrons getting more energy than ions.
Our conclusion is thus that overall, ions and electrons are almost equally energized.
It is, however, essential to know whether this remains true at realistic mass ratios. 
With no guide field, our simulations with $m_\ion/m_\lec = 12$ and $m_\ion/m_\lec=50$ show a variation of $\sim 3\%$, 
which cannot be distinguished 
from the variation due to the different background magnetizations of these simulations. 
Consequently, and even if higher mass ratios should be tested, it seems that the $\sim60\%$/$40\%$
repartition will hold at larger $m_\ion/m_\lec$.
The case with a guide magnetic field will be explored in more depths in a future work.

The maximal Lorentz factor of the background particles (ii) is shown in Fig.~\ref{fig:max_Lorentz_factor}. 
A larger guide field leads to lower highest Lorentz factors, 
and this is expected because the reconnection electric field becomes weaker 
with increasing guide field \citep{Melzani2014a}.
A log-log plot shows that the rate of increase is roughly $\gamma_\mathrm{max} \propto t^s$, 
with $s$ ranging from 0.7 to 1.1 as the magnetization rises, with identical values for ions and electrons.
Particles inside the islands follow the same trend.
We stress that this is faster than 
in collisionless shock acceleration where $\gamma_\mathrm{max} \propto t^{0.5}$ \citep{Bell1978}.
We also remark that the maximal Lorentz factor does not saturate.
It should saturate in very large systems when the inter-island distance 
becomes larger than the particle Larmor radii in the reconnected field $B_x$,
and when island merging ceases.

\begin{table*}[tbp]
\caption{\label{tab:energy_distribution}Energy distribution between fields and particles at the end of different simulations.
cs stands for the current sheet population, bg for the background population.
The final energy in the electric field is $E_\tot$, and is dominated by the energy in $E_y$.
The energy in $B_x$ is denoted by $B_x$.
As explained in Fig.~\ref{fig:energy}, 
$\mathcal{E}_0$ is the total (mostly magnetic) initial energy in
the ``reconnection area'', which is the area from where the particles reach the current sheet before 
the end of the simulation. 
Because of the difficulty to precisely measure this area, the percentage are to be taken 
with a $\pm 5$\% relative uncertainty. The energy repartition between ions and electrons is not affected by this.
}
\centering
\begin{tabular}{c|c|cc|c||c|c|c||cc|cc}
 \multicolumn{5}{c||}{simulation parameters} & \multicolumn{3}{c||}{final energy in \% of $\mathcal{E}_0$} & \multicolumn{4}{c}{ion/electron energy repartition} \\
 $\wce/\wpe$ & $n_\bg/n_\cs(0)$ & $\sigma^\mathrm{hot}_{\ion}$ & $\sigma^\mathrm{hot}_{\lec}$ & $V^\mathrm{R}_\mathrm{A,in}$ & $B_x$ & $E_\tot$ & kinetic energy & cs ions & cs elec & bg ions & bg elec \\
   &     &     &     &      &      &     &      &      &      &      &      \\ 
\hline
\multicolumn{2}{l|}{$m_\ion/m_\lec = 50$} & & & & & & & & & & \\
 6 & 0.1 & 7.1 & 260 & 0.93 &   21\% &  5\% &   74\% &   54\% &   46\% &   63\% &   37\% \\
 \hline
\multicolumn{2}{l|}{$m_\ion/m_\lec = 25$} & & & & & & & & & & \\ 
 1 & 0.1 & 0.4 & 9.9 & 0.53 & 11\% & 1\% & 48\% & 67\% & 33\% & 60\% & 40\% \\
 3 & 0.1 & 3.6 & 83  & 0.88 & 20\% & 3\% & 62\% & 56\% & 44\% & 63\% & 37\% \\
 3 & 0.3 & 1.2 & 27  & 0.73 & 17\% & 4\% & 74\% & 61\% & 39\% & 62\% & 38\% \\
 6 & 0.1 &  14 & 260 & 0.97 & 18\% & 4\% & 73\% & 53\% & 47\% & 60\% & 40\% \\
\hline
\multicolumn{2}{l|}{$m_\ion/m_\lec = 12$} & & & & & & & & & & \\
 3 & 0.1 & 7.5 & 83  & 0.93 & 21\% & 2\% & 72\% & 55\% & 45\% & 60\% & 40\% \\
 \hline
\multicolumn{2}{l|}{$m_\ion/m_\lec = 1$} & & & & & & & & & & \\
 3 & 0.1 & 83 & 83  & 0.99 & 34\% & 8\% & 87\% & 50\% & 50\% & 50\% & 50\% \\
\end{tabular}
\end{table*}

\begin{table*}[tbp]
\caption{\label{tab:energy_distribution_guidefield}Same as Table~\ref{tab:energy_distribution},
but for cases where there is a guide field. Here $m_\ion/m_\lec=25$.
In this case $\mathcal{E}_0$ does not include the guide magnetic field $B_y$, 
because this component is not transferred to the particles (the energy in $B_y$ remains constant 
to within 5\% throughout the simulation).
}
\centering
\begin{tabular}{c|c|cc|c||c|c|c||cc|cc}
 \multicolumn{5}{c||}{simulation parameters} & \multicolumn{3}{c||}{final energy in \% of $\mathcal{E}_0$} & \multicolumn{4}{c}{ion/electron energy repartition} \\
 $\wce/\wpe$ & $n_\bg/n_\cs(0)$ & $\sigma^\mathrm{hot}_{\ion}$ & $\sigma^\mathrm{hot}_{\lec}$ & $V^\mathrm{R}_\mathrm{A,in}$ & $B_x$ & $E_\tot$ & kinetic energy & cs ions & cs elec & bg ions & bg elec \\
   &     &     &     &      &      &     &      &      &      &      &      \\ 
\hline
\multicolumn{2}{l|}{$B_\mathrm{G}/B_0=0.5$} & & & & & & & & & & \\
 3 & 0.1 & 3.6 & 89  & 0.81 & 16\% & 4\% & 43\% & 60\% & 40\% & 46\% & 54\% \\
 \hline
\multicolumn{2}{l|}{$B_\mathrm{G}/B_0=1$} & & & & & & & & & & \\
 3 & 0.1 & 3.6 & 89  & 0.66 & 18\% & 5\% & 45\% & 57\% & 43\% & 33\% & 67\% \\
\end{tabular}
\end{table*}


\subsection{Discussion}

\subsubsection{Acceleration mechanisms}
The main acceleration mechanism for the background population (ii) is direct acceleration by the reconnection electric field.
However, other acceleration mechanisms are present. 
It can be seen by dividing the background population into several subgroups, that comprise particles in a slab $x_\mathrm{min}<x<x_\mathrm{max}$ at $t=0$.
We can then follow the spatial evolution of these slabs, along with the evolution of their momentum distribution function.
After gaining energy in the acceleration region by $E_\mathrm{rec}$, 
background particles escape along the reconnection exhausts, and circle at the island periphery following the strong magnetic field structure.
Contraction of the islands (when they merge) creates strong motional electric fields $\b{E} = -\b{v}\wedge\b{B}$ that accelerate these particles.
Also, two merging islands create a reconnection event with a
reconnection electric field along $+\hat{\b{y}}$, that can transfer energy to particles.
The combination of these three mechanisms is also reported for pairs 
by \citet{Bessho2012} and \citet{Sironi2014}.

We find no trace of Fermi acceleration between the converging inflows, as studied analytically or with test particles by
\citet{Giannios2010}, \citet{Kowal2011}, or \citet{Drury2012},
because particles cannot cross the current sheet \citep[see][]{Melzani2014a}, but bounce back and forth only inside the diffusion region
where they are constantly accelerated by the reconnection electric field.
Fermi acceleration is possible only if inflowing particles of species $s$ are energetic enough to cross the diffusion region,
i.e., have a Larmor radius $\gamma m_s v / eB_0$ larger than the diffusion zone width $\delta_s$. 
The latter is roughly one inertial length measured at its center \citep{Melzani2014a}, so that the crossing condition can be written 
$\gamma v/c > (\sigma^\mathrm{cold}_s n_\bg/n_\cs)^{1/2}$ (with $\sigma^\mathrm{cold}_s$ the background magnetization defined by 
Eq.~\ref{equ:acc_sigma_s_cold}, and $n_\bg$ and $n_\cs$ the background and current sheet density). 
The right hand side is larger than unity for a relativistic inflow magnetization. 
This mechanism thus requires
already accelerated particles in the inflow, 
which is possible for very high temperatures, 
or for an out-of-equilibrium component 
pre-accelerated by other mechanisms outside of the diffusion region such as neighboring reconnection sites, 
or large scale flow turbulence (see Sect.~\ref{sec:other_acc_mechanism}). 
We have neither of this in our simulations, and their presence in real situations 
should be addressed.

We now turn to the building of the power-law spectrum in our simulations.
The basic idea is that particles enter the acceleration region at all distances from the central X-point,
with those entering near the center being deviated toward the exhausts by $B_x$ more slowly than those entering at the edges.
They thus gain more energy, and a monoenergetic inflow is transformed into 
a broader distribution. 
The ingredient to build a power-law, underlined by \citet{Zenitani2001}, 
is that relativistic particles have a Larmor radius increasing with their Lorentz factor: 
high-energy particles rotate slowly in $B_x$ and are held in the acceleration region even longer as they are accelerated, 
thus facilitating the creation of hard tails. 
With this argument alone, \citet{Zenitani2001} predict a power-law with index $p\propto cB_x/E_\mathrm{rec}$, but their model is 
very simple. \citet{Bessho2012} derive analytically the spectra of particles accelerated by $E_\mathrm{rec}$ 
and escaping from the X-point, and find 
$\dif n/\dif\gamma \propto \gamma^{-1/4}\exp\{-a\gamma^{1/2}\}$. It is a power-law with an exponential cutoff, 
occurring at higher energies for relativistic X-points because $a\propto (cB_x)/E_\mathrm{rec}$. 
In the end, the X-point accelerated particles gain further energy around the 
contracting islands and during reconnection events between merging islands, 
to result in the distribution that we show in this paper.

\subsubsection{Condition for hard high-energy tails}\label{sec:discussion_cond_high_energy}

It appears from our data that the hardness of the energy distribution for species $s$ is controlled 
by its background magnetization $\sigma_s^\mathrm{hot}$ and by the Alfv\'en speed $V^\mathrm{R}_\mathrm{A,in}$.
The power-law is harder for larger magnetizations (Table~\ref{tab:param_physical_tearing}).
For a fixed magnetization $\sigma_\lec^\mathrm{hot}=83$, the simulations with $m_\ion/m_\lec = 1$, 12, and 25,
feature different inflow Alfv\'en speeds $V^\mathrm{R}_\mathrm{A,in}=0.988$, 0.93, and 0.88, 
and also different power-law indexes for electrons: $p=1.5$, 1.6, 1.8, respectively.
Similarly, when compared to our work at a given electron magnetization, 
the PIC simulations of \citet{Sironi2014} for pair plasmas present harder distributions, simply because 
with pairs and a given $\sigma_\lec$, the Alfv\'en speed is larger.
A larger inflow Alfv\'en speed thus leads to harder distributions.

This can be interpreted as follows.
The building of a high-energy tail for species $s$ requires two ingredients: 
an inflow magnetization $\sigma_s > 1$ so that the transfer of magnetic energy
can exceed the particles rest-mass, and thus accelerate them to relativistic energies; 
and a large enough ratio $E_\mathrm{rec}/(cB_0)$ in order to have a hard distribution. 
The latter condition roughly states that the residence time $\gamma m_s/(q_sB_x)$
of a particle in the acceleration region
must be large compared to its acceleration time $\gamma m_sc/(q_sE_\mathrm{rec})$ by $E_\mathrm{rec}$.
This is supported by the analytical models of \citet{Zenitani2001} and \citet{Bessho2012} previously cited. 
Given that the reconnection rate\footnote{The residence time $\gamma m_s/(q_sB_x)$ implies the reconnected field $B_x$,
and not the inflowing component $B_0$ as in the reconnection rate. 
However, the ratio $B_0/B_x$ is roughly the same for all magnetizations and all simulations.}
$E^* = E_\mathrm{rec}/(B_0 V^\mathrm{R}_\mathrm{A,in}\cos\theta)$ 
lies in a narrow range for various simulations 
\citep[$E^*\sim 0.14$-0.30,][]{Melzani2014a}, 
the condition of a large ratio $E_\mathrm{rec}/(cB_0) = E^* V^\mathrm{R}_\mathrm{A,in}\cos\theta /c$ consequently 
translates into a 
relativistic inflow Alfv\'en speed $V^\mathrm{R}_\mathrm{A,in}$
and a not too strong guide field ($\cos\theta=(1+B^2_\mathrm{G}/B^2_0)^{-1/2}$).

For the ions, a hard distribution requires $\sigma_\ion^\mathrm{hot} \gg 1$ and $V^\mathrm{R}_\mathrm{A,in} \sim c$.
But the condition $V^\mathrm{R}_\mathrm{A,in} \sim c$ is equivalent to a total magnetization 
$\sigma_{\ion+\lec}(B_\tot) > 1$ (Eq.~\ref{equ:def_Alfven_speed}),
which is fulfilled already if $\sigma_\ion^\mathrm{hot} \gg 1$.
For the electrons, a hard distribution requires $\sigma_\lec^\mathrm{hot} \gg 1$ 
and $V^\mathrm{R}_\mathrm{A,in} \sim c$.
The latter translates to $\sigma_{\ion+\lec}(B_\tot) > 1$, so that we have
(if we neglect temperature effects): 
$\sigma_\lec^\mathrm{cold} \sim (1+m_\ion/m_\lec)\sigma_{\ion+\lec} > 1+m_\ion/m_\lec \sim 2000$.
Here the condition on the Alfv\'en speed 
is consequently more stringent than that on the magnetization.
We conclude that hard ion distributions are obtained when $\sigma_\ion^\mathrm{hot} \gg 1$,
and hard electron distributions when $V^\mathrm{R}_\mathrm{A,in} \sim c$.

We finally remark that the Alfv\'en speed, and thus also the ratio $E_\mathrm{rec}/B_0$,
saturates at $c$ for large magnetizations.
We consequently expect the particle distribution hardness to also saturate to some value
with, as indicated by our simulations, a power-law index $p$ slightly below $1.2$.
Reconnection in environments with $\sigma_\ion^\mathrm{hot} \gg 1$ should thus produce 
ion and electron distributions with power-law index $p\lesssim1.2$.

\subsection{Astrophysical outlook}
\label{sec:astro_outlook}

\begin{table*}[tbp]
\caption{\label{tab:real_parameters}Order of magnitude for physical parameters in astrophysical environments.
}
\centering
\begin{tabular}{l|c|c|c|c|c}
 Objects with ion-electron plasmas (with also pairs) & $B$ (G) & $n_\lec$ ($\mathrm{cm^{-3}}$) & $T_\lec$ (K) & $\sigma^\mathrm{cold}_{\lec}$ & $V^\mathrm{R}_\mathrm{A,in}/c$ \\
\hline
 Microquasar coronae, X-ray emitting region\tablefootmark{a}               & $10^5$-$10^7$       & $10^{13}$-$10^{16}$  & $10^9$         & $10^{-1}$-$10^5$& 0.003-1 \\
 AGN coronae, X-ray emitting region\tablefootmark{b}                       &                     &                      & $10^9$         & 1.7-180         & 0.03-$0.3$ \\
 Giant radio galaxy lobes\tablefootmark{c}                                 & $10^{-6}$-$10^{-5}$ & $3\times10^{-6}$     & $10^6$         & 0.8-80          & 0.02-$0.2$ \\
 Extragalactic jet, $\gamma$-ray emitting region ($<0.05$\,pc)\tablefootmark{d} & 12             & 80                   &                & $2\times10^5$   & $\sim 1$ \\
 Extragalactic jet, radio emitting region (kpc scales)\tablefootmark{e}    & 1-$3\times10^{-5}$  & 0.8-$5\times10^{-8}$ &                & 500-2500        & $\sim 1$ \\
 GRB jet, at radius of fast reconnection\tablefootmark{f}                  & $7\times10^8$       & $10^{10}$            & $10^8$         & $5\times10^{12}$& $0.9$ \\
 & & & & & \\
 Objects with pair plasmas & $B$ (G) & $n_\lec$ ($\mathrm{cm^{-3}}$) & $T_\lec$ (K) & $\sigma^\mathrm{cold}_{\lec}$ & $V^\mathrm{R}_\mathrm{A,in}/c$ \\
\hline
 At the termination shock of pulsar winds\tablefootmark{g}                 & $10^4$              & 0.1-10               &                & $10^{13}$      & $\sim 1$ \\
 In a pulsar wind nebulae\tablefootmark{h}                                 & $5\times10^{-3}$    & 5 to $10^3$          & $\gamma\sim 10$-$10^9$& $<0.5$    & $0.6$ \\
\end{tabular}
\tablefoot{
\tablefoottext{a}{Analytical disk and corona models, \citet{deGouveia2005,Matteo1998,Merloni2001,Reis2013}; matching observed spectra with radiation models, \citet{DelSanto2013,Romero2014}.}
\tablefoottext{b}{Analytical disk and corona models, \citet{Matteo1998,Merloni2001,Reis2013}.}
\tablefoottext{c}{Observations, \citet{Kronberg2004}.}
\tablefoottext{d}{Analytical model assuming $\sigma_\ion^\mathrm{cold}=100$, \citet{Giannios2009}. See also \citet{Giroletti2004} for magnetic field measurements (0.2\,G, but on larger scales).}
\tablefoottext{e}{Observations, \citet{Schwartz2006}. See also \citet{Romanova1992}.}
\tablefoottext{f}{Analytical model, \citet{McKinney2012b}. Pairs are also present, with $n_\mathrm{pair}\sim10n_\lec$.}
\tablefoottext{g}{Analytical model and observations, \citet{Bucciantini2011,Sironi2011b}.}
\tablefoottext{h}{Analytical model and observations, \citet{Atoyan1996,Meyer2010,Uzdensky2011,Cerutti2013}. The plasma distribution function is a broken power-law with Lorentz factors $\gamma$ in the indicated range. We note that \citet{Cerutti2013} considers only the high-energy electron population, and hence has larger magnetizations.}
}
\end{table*}

\subsubsection{Objects and orders of magnitude}

Magnetic reconnection can play a major role for four particular purposes:
high-energy flare production, steady emission of radiation, large scale outflow launching, and plasma heating
or non-thermal particle production.
We discuss in which objects these phenomena are observed, give order of magnitudes for the main parameters, 
and point out where our work is applicable. 
Table~\ref{tab:real_parameters} summarizes the physical conditions encountered in the objects mentioned here.

Flare-like emission of high-energy photons is observed 
for example in the $\gamma$-ray region of AGN jets \citep{Giannios2010}, or
in microquasar and AGN coronae \citep{Matteo1998,Merloni2001,Reis2013}.
Concerning AGN jets, flares in the GeV-TeV range are observed, and may come from close to the AGN 
\citep[$<0.05$\,pc,][]{Giroletti2004}.
There, \citet{Giannios2009} assume an ion magnetization $\sigma_\ion^\mathrm{cold} \sim 100$, which gives 
an Alfv\'en speed $\sim c$ (Table~\ref{tab:real_parameters}, line~d).
Concerning the coronae of AGNs and microquasars, various models constrained by observations predict 
ion magnetizations in the range $\sigma_\ion^\mathrm{cold} \sim 10^{-5}$-$10^2$ 
(Table~\ref{tab:real_parameters}, lines~a and~b). In the most magnetized case, the Alfv\'en speed is $\sim c$.
According to our results, reconnection in such environments should produce 
electron distributions with hard tails ($p\sim 1$), and 
if we apply our results for electrons to the ions, the latter should also 
form hard-tails ($p\sim 2$).
Quite generally, high-energy flares can be explained by reconnecting events under three conditions: 
the large scale magnetic field must possess enough energy and the large scale flow 
or the flow turbulence must lead to enough reconnection events; 
the reconnection process must produce high-energy particles with hard distributions;
these high-energy particles must be able to radiate.
The first point is linked to the large scale properties of the object and can be investigated with simulations on large scales, 
the last two points concern microphysics and must be addressed with first principle simulations.
We believe to have answered the second point: magnetic reconnection does produce hard high-energy distributions
whenever the ion magnetization is above unity, which can indeed be the case in the environments mentioned above.
Concerning photon emission, 
we expect particles trapped inside the islands (population i) to produce mostly thermal synchrotron-Bremsstrahlung.
On the other hand, particles accelerated at the X-point (population ii) are likely to 
radiate collimated beams when suddenly encountering the strong magnetic field structure at the end of the exhausts,
at the island entrance.
This was demonstrated in the no guide field case by \citet{Cerutti2012b} in pair plasmas, and should also hold 
for ion-electron plasmas because the overall magnetic structure is not too different.
The radiation spectrum is then anisotropic, and reaches energies of the order of $\gamma^2 \wce$ with 
$\gamma$ the particles Lorentz factor and $\wce = eB/m_\lec$.
With a guide field, radiation should occur together with particle acceleration in the $\b{E}\cdot\b{B}\neq 0$ area,
because particles then swirl around the guide field.
It should consequently be more regular and less flare-like \citep{Cerutti2013}. 
Since reconnection with a guide field is more generic in the complex magnetic field structures of 
coronae or jets than antiparallel reconnection, 
the question of photon emission in such a case is very relevant.

Magnetic reconnection can produce radiation in a large range of frequency, which for the synchrotron component depends on the 
strength of the reconnecting magnetic field.
An example is the radio emission, on kilo-parsec scales, of extragalactic jets. Explanation of these spectra 
can invoke magnetic reconnection events, in particular to explain the hard photon indexes \citep{Romanova1992}.
Observations indicate electron magnetizations in the range $\sigma_\lec^\mathrm{cold} \sim 500$-2500 
(Table~\ref{tab:real_parameters}, line~e), 
which corresponds to ion magnetizations $\sigma_\ion^\mathrm{cold} \sim 0.3$-1.3 and to Alfv\'en speeds $\sim 0.5$-$0.8c$.
This is between the two cases $\wce/\wpe=1$ and $\wce/\wpe=3$ of our study, for which the kinetic energy takes 
a large amount of the magnetic energy 
(between 45\% and 60\%, Tables~\ref{tab:energy_distribution} and \ref{tab:energy_distribution_guidefield}),
and is distributed as 60/40\% to 30/70\% between, respectively, ions and electrons, depending on the guide field strength.
Most importantly, for these parameters the background accelerated electrons form power-laws with 
indexes between 4.5 and 1.5 (Table~\ref{tab:param_physical_tearing} and Fig.~\ref{fig:histo_lec_summary_all}), 
which can then indeed emit hard spectra.

Large scale magnetic reconnection events may be at the origin of 
large scale transient jets in microquasar systems \citep{deGouveia2005,deGouveia2010,Kowal2011,McKinney2012,Dexter2013},
for example via a magnetic field reversal accreted onto a magnetically arrested disk.
Reconnection then occurs in the accretion disk coronae, near the black hole, where particle densities and 
magnetic fields are high. 
\citet{deGouveia2005} estimate the ion magnetization close to the black hole to be $\sigma_\ion^\mathrm{cold} \sim 60$,
i.e., $\sigma_\lec^\mathrm{cold} \sim 10^5$ 
(Table~\ref{tab:real_parameters}, line~a). 

Magnetic reconnection can also efficiently heat the plasma by dissipating the magnetic field energy.
This is invoked for the heating of AGNs and microquasar coronae, that must reach 
electron temperatures of the order of $10^9$\,K to be able to inverse-Compton seed photons to X-ray energies.
In microquasars, a coronal population of non-thermal high-energy electrons is also required by the observation 
of MeV photons \citep{Poutanen2014}.
Alfv\'en speeds in these coronae are estimated to be of the order of $0.003c$-$c$ (Table~\ref{tab:real_parameters}, line~a).
In our corresponding simulations, the final kinetic energy is a large fraction of 
the initial magnetic energy
(48\% at $V_\mathrm{A}=0.5c$, up to 75\% at $V_\mathrm{A}=0.88c$, Table~\ref{tab:energy_distribution}), 
thus allowing an efficient energy transfer to the plasma.
For this range of Alfv\'en speeds, accelerated electron distributions can be steep ($p>4$ for $V_\mathrm{A}=0.5c$)
or hard ($p\sim 1.2$ for $V_\mathrm{A}=0.97c$), and in the latter case reconnection can indeed produce a non-thermal population.
An important point is the question of the energy repartition between ions and electrons. 
Our results show that energy 
is almost equally distributed between ions (60\% to 30\%) and electrons (40\% to 70\%), so that we do not expect 
large temperature differences \citep[as studied in some models,][]{Matteo1997} from this heating mechanism.
Other models of the MeV component from microquasars invoke 
the emission of electrons in the jet, and require hard electron distributions with indexes $p\sim 1.5$ \citep{Zdziarski2014}.
The ion magnetization in these models needs to be larger than unity. Our work demonstrates that magnetic reconnection
in these conditions can provide such hard electron spectra.
Also, very similar conditions are expected in the lobes of radio galaxies 
\citep[$V_\mathrm{A}\sim 0.02$-$0.2c$,][]{Kronberg2004}, and our conclusions also apply there,
especially for the ion/electron energy repartition.

A final application concerns the extraterrestrial PeV neutrinos detected by IceCube \citep{IceCube2013}.
They can come from the photopion ($p\gamma$) interaction of high-energy protons or ions produced by high-energy machines 
\citep[such as, e.g., GRBs,][]{Petropoulou2014}.
The ability of magnetic reconnection to accelerate ions in highly magnetized environments is then very relevant.
We find that the highest Lorentz factor for ions follows the same trend as that for electrons ($\gamma\propto t^s$, $s\sim 0.7\text{-}1.1$),
and for ion magnetization $\sigma_\ion \gg 1$ we expect ions to feature the same power-law spectra as electrons,
with a slope $p \lesssim 1.2$.

\subsubsection{Scaling of the results, importance of radiative braking, Compton drag and pairs}
The reconnection physics with the Harris geometry described in this paper depends only on 
the inflow plasma magnetization and temperature, and not on the absolute values of magnetic field
and particle number density. For example, reconnection in a microquasar corona close to the hole, and in 
the $\gamma$-ray emitting region of an extragalactic jet, 
takes place with the same magnetizations (Table~\ref{tab:real_parameters})
and thus feature the same reconnection rate, particle spectra or energy repartition, even if 
magnetic field strengths differ by six orders of magnitude
(provided, however, that the geometry is the same).
This is true as long as effects such as radiative braking by emission or Compton drag, or pair annihilations, 
do not perturb the reconnection physics.
Such effects imply the actual values of magnetic field and particle number densities, and can lead to a very different
physics for a same magnetization.
In order to evaluate this, we estimate in Appendix~\ref{sec:appendix_radiation} when an electron looses 
a significant fraction of its energy during a time or over a length scale 
dynamically important for the magnetic reconnection physics.
These time and length scales are taken as a cyclotron or plasma period, or as an inertial scale.
Particles will eventually radiate and cool further away, but with no influence on the reconnection physics.
We summarize our conclusions here.

Radiative braking due to synchrotron radiation remains negligible on cyclotron scales
as long as $(\gamma/100)^2\,B < 10^{11}$\,G (Eq.~\ref{equ:sync_emmision_braking}), 
where $\gamma$ is the Lorentz factor of an electron.
This is negligible for all objects of Table~\ref{tab:real_parameters},
except in the pulsar wind nebulae where $\gamma$ can reach $10^9$ \citep{Meyer2010}.
Radiative braking due to Coulomb collisions can be estimated by assuming a thermal Bremsstrahlung,
and remains negligible on inertial length scales as long as 
$(T_\lec/m_\lec c^2)^{1/2} (n_\lec/5\times10^{12}\mathrm{cm}^{-3})^{3/2} < 1$ (Eq.~\ref{equ:Brem_emmision_braking}).
This is the case for objects of Table~\ref{tab:real_parameters}, except in microquasar coronae close to the black hole
where $n_\lec$ is high.

Compton drag does not affect the electron dynamics on inertial length scales 
as long as $(\gamma/100)(1\mathrm{cm^{-3}}/n_\lec)^{1/2}\,U_\mathrm{ph} < 10^{10}\mathrm{erg/cm^3}$,
with $U_\mathrm{ph}$ the radiation field energy density (Eq.~\ref{equ:IC_loss}).
For a blackbody spectrum, this remains true for photon temperatures $T_\mathrm{ph} < 10^6$\,K.
Consequently, objects with electron temperatures $T_\lec > 10^6$\,K on scales large enough so that 
the optical depth is important and photons are thermalized, are in the range where Compton drag is efficient.
It should be noted that the electrons locally heated by the magnetic reconnection cannot thermalize the 
radiation, because the reconnection region is optically thin (Eq.~\ref{equ:opacity_Compton}).
The photon energy density $U_\mathrm{ph}$ produced by the magnetic reconnection 
is then to be computed from the synchrotron or Bremsstrahlung emissivity, assuming that 
the emission takes place over a volume $(ad_\lec)^3$ with $d_\lec$ the electron inertial length and $a$ a geometrical factor.
Compton drag against the synchrotron photons is negligible as long as 
$a (B/1\mathrm{G})^2 (\gamma/100)^3 < 10^{36}$ (Eq.~\ref{equ:sync_compton_drag}),
and that due to Bremsstrahlung emission as long as 
$a (T_\lec/10^8\mathrm{K})^{1/2} (n_\lec/1\mathrm{cm^{-3}}) (\gamma/100) < 10^{38}$ (Eq.~\ref{equ:Brem_compton_drag}).
Compton drag by these photon fields is thus negligible for all objects of Table~\ref{tab:real_parameters}.

Finally, photons of energy above $m_\lec^2 c^4 / \epsilon_0$ can annihilate with ambient photons 
(of typical energy $\epsilon_0$) to produce pairs.
This can be the case if $B (\gamma/100)^2 > 2\times10^9$\,G for high-energy synchrotron photons, 
or if $T_\lec > 10^9$\,K for Bremsstrahlung (Eqs.~\ref{equ:hnu_sync} and~\ref{equ:hnu_Brem}).
Inverse Compton events can also produce such photons if the electron Lorentz factors are $\gamma > m_\lec c^2 / \epsilon_0$
(see Appendix~\ref{sec:appendix_radiation}).

In any case, pair creation will disturb the reconnection dynamic only 
if the creation occurs inside or close to the reconnection region.
The mean-free-path $l_{\gamma\gamma}$ 
of high-energy photons should thus be compared to an inertial length $d_\lec$.
For a blackbody gas of photons at temperature $T_\mathrm{ph}$, we have 
$l_{\gamma\gamma,\mathrm{BB}}/d_\lec = (1\mathrm{cm}^{-3}/n_\lec)^{1/2} (10^6\mathrm{K}/T_\mathrm{ph})^3$
(Eq.~\ref{equ:gammagamma_annihilation_CN}).
However, as underlined previously, a blackbody spectrum of photons is not easy to achieve. 
If photons cannot be thermalized, then the $\gamma\gamma$ opacity must be
computed from the rate of production of photons by synchrotron and Bremsstrahlung radiation in the reconnection region. 
Concerning Bremsstrahlung emission, we find with Eq.~\ref{equ:gammagamma_annihilation_Brem}
that pairs form far away from the reconnection region for all objects of Table~\ref{tab:real_parameters}.
Concerning synchrotron emission, we find with Eq.~\ref{equ:gammagamma_annihilation_sync}
that for radio lobes, radio emitting regions of extragalactic jets, or pulsar wind nebulae,
$l_{\gamma\gamma,\mathrm{sync}}\gg d_\lec$ holds, so that pairs form far away from the reconnection site; 
while for microquasar coronae close to the hole, for extragalactic jet $\gamma$-ray regions,
for GRB jets, or for pulsar wind termination shocks, we have 
$l_{\gamma\gamma,\mathrm{sync}} \ll d_\lec$ and pairs form inside the reconnection region.

\subsubsection{Other acceleration sites during reconnection}\label{sec:other_acc_mechanism}
In this manuscript, we investigate particle acceleration in or close to the diffusion region.
Other energy conversion locations exist around reconnection sites.
One is along the magnetic separatrices far downstream of the diffusion region 
\citep[observed, e.g., at the magnetopause: ][]{Khotyaintsev2005},
on length scales of hundreds of ion inertial lengths. 
There, magnetic energy conversion occurs
as the plasma flows 
through the complex structure of collisionless non-linear waves
(slow shock(s), compound wave, rotational wave). 
Instabilities and parallel electric fields 
in these regions can produce thermal and non-thermal electrons 
\citep{Drake2005,Egedal2009,Egedal2012}.
This shock structure has been investigated in the non-relativistic case \citep{Liu2012,Higashimori2012}. 
In a relativistic situation, the different phase speeds of the waves may lead to different results.
Of particular interest is the energy distribution between bulk, thermal, ion, and electron components,
and its importance relative to the locations discussed here.

Another site for particle acceleration is at the dipolarization front \citep{Vapirev2013}, where the first reconnected 
field lines are swept away and drag the ambient plasma. Such a situation is prone to instabilities and particle acceleration.

Also, turbulence associated with magnetic reconnection can lead to particle acceleration via a second order Fermi process.

We emphasize, however, that high-energy particle production in and near the diffusion zone, directly by the reconnection electric field
as discussed in the present manuscript, 
should be of great importance for relativistic inflow magnetizations because the electric field is very large
\citep[$E_\mathrm{rec}/B_0 \sim 0.2V_\mathrm{A,in}^\mathrm{R} \sim 0.2c$, see ][]{Melzani2014a}.

\subsubsection{Open questions}
The present study brings useful insights on the properties of magnetic reconnection, but remains simplified in many respects.
Reconnection configurations in real environments are likely to often involve guide fields, 
but also asymmetric plasmas and fields from each side of the current sheet \citep{Aunai2013,Eastwood2013},
or normal magnetic fields (i.e., along $\hat{\b{x}}$ here) due to the ambient field \citep[e.g., for the magnetotail,][]{Pritchett2005b}.
Magnetic reconnection is also likely to be forced by external plasma and field line motions. 
The reconnection electric field is then imposed by the forcing \citep{Pei2001,Pritchett2005,Ohtani2009,Klimas2010},
and can be larger than in the spontaneous case. Particle acceleration can consequently be enhanced.
The initial equilibrium can also have an impact on the late evolution and particle distributions. 
Study of 2D situations such as X-point collapse or force-free equilibrium 
\citep[e.g.,][]{Pahlen2013,Liu2014} show little differences with the Harris case, and we expect them to produce the same kind of distributions.
However, full 3D initial configurations can lead to very different outcomes, as suggested by the few existing kinetic studies
\citep{Baumann2012,Olshevsky2013}. 
Reconnection and particle acceleration at 3D nulls or at quasi separatrix layers \citep{Pontin2011} deserves further research.

A crucial question concerns the validity of our results in a real 3D reconnection event. 
Magnetic islands then become extended filaments, modulated or broken by instabilities in the third dimension or 
by a lack of coherence of the tearing instability \citep{Daughton2011,Kagan2012,Markidis2013}.
For this reason we may expect more particle mixing,
but current sheet particles may also still be trapped in the strong magnetic structure surrounding the filaments.
Particle acceleration at X-points may also be disturbed.
However, first 3D results in pair plasmas by \citet{Sironi2014} are encouraging 
in showing that energization is still efficient, and leads to 
power-law tails with similar indexes, essentially because the small scale physics around the X-point 
and during filament mergings is the same as in 2D. 

There is also a strong need to better understand the interplay between large
and small scales. Coronal heating by reconnection, or large scale outflow launching, 
are cases where the large scale flow sets the conditions for the occurrence of reconnection, 
which in turn largely modifies the large scale flow conditions.
For example, \citet{Jiang2014} show that global simulations of 
the formation of an accretion disk corona requires an understanding of the role of reconnection in 
the MRI turbulence. 
Shocks, possibly collisionless, are also fundamental micro-physical processes 
that shape the flow on all scales of accreting black holes \citep{Walder2014}.
\citet{Daldorff2014} illustrate
the power of a coupled MHD/PIC approach with a simulation of the Earth magnetosphere.

Lastly, we emphasize that the ability of magnetic reconnection to accelerate protons or heavier ions 
is a key question. First because they can produce mesons and then pairs, which can lead to a different 
photon spectrum. 
Second because this channel can produce neutrinos, and characterizing the neutrino spectrum 
from high-energy objects is compulsory to distinguish it from those predicted by dark matter models.
With high-energy extraterrestrial neutrinos being now detected \citep{IceCube2013}, this is a very 
exciting perspective.

\begin{acknowledgements}
  This work was performed using HPC ressources from GENCI-CCRT/IDRIS/CINES/TGCC (grant x2013046960).
  Tests were conducted at p\^ole scientifique de mod\'elisation num\'erique,
  PSMN, at the ENS Lyon.
  We acknowledge the Programme National Hautes Energies (PNHE) for financial support.
\end{acknowledgements}
\bibliographystyle{apj} 
\bibliography{biblio_particles.bib}

\begin{thebibliography}{83}
\expandafter\ifx\csname natexlab\endcsname\relax\def\natexlab#1{#1}\fi

\bibitem[{{Atoyan} \& {Aharonian}(1996)}]{Atoyan1996}
{Atoyan}, A.~M. \& {Aharonian}, F.~A. 1996, \aaps, 120, C453

\bibitem[{{Aunai} {et~al.}(2013){Aunai}, {Hesse}, {Zenitani}, {Kuznetsova},
  {Black}, {Evans}, \& {Smets}}]{Aunai2013}
{Aunai}, N., {Hesse}, M., {Zenitani}, S., {Kuznetsova}, M., {Black}, C.,
  {Evans}, R., \& {Smets}, R. 2013, Physics of Plasmas, 20, 022902

\bibitem[{{Baumann} \& {Nordlund}(2012)}]{Baumann2012}
{Baumann}, G. \& {Nordlund}, {\AA}. 2012, \apjl, 759, L9

\bibitem[{{Bell}(1978)}]{Bell1978}
{Bell}, A.~R. 1978, \mnras, 182, 147

\bibitem[{Bessho \& Bhattacharjee(2012)}]{Bessho2012}
Bessho, N. \& Bhattacharjee, A. 2012, The Astrophysical Journal, 750, 129

\bibitem[{{Birk} {et~al.}(2001){Birk}, {Crusius-W{\"a}tzel}, \&
  {Lesch}}]{Birk2001}
{Birk}, G.~T., {Crusius-W{\"a}tzel}, A.~R., \& {Lesch}, H. 2001, \apj, 559, 96

\bibitem[{{Bucciantini} {et~al.}(2011){Bucciantini}, {Arons}, \&
  {Amato}}]{Bucciantini2011}
{Bucciantini}, N., {Arons}, J., \& {Amato}, E. 2011, \mnras, 410, 381

\bibitem[{{Cerutti} {et~al.}(2012{\natexlab{a}}){Cerutti}, {Uzdensky}, \&
  {Begelman}}]{Cerutti2012}
{Cerutti}, B., {Uzdensky}, D.~A., \& {Begelman}, M.~C. 2012{\natexlab{a}},
  \apj, 746, 148

\bibitem[{{Cerutti} {et~al.}(2012{\natexlab{b}}){Cerutti}, {Werner},
  {Uzdensky}, \& {Begelman}}]{Cerutti2012b}
{Cerutti}, B., {Werner}, G.~R., {Uzdensky}, D.~A., \& {Begelman}, M.~C.
  2012{\natexlab{b}}, \apjl, 754, L33

\bibitem[{{Cerutti} {et~al.}(2013){Cerutti}, {Werner}, {Uzdensky}, \&
  {Begelman}}]{Cerutti2013}
---. 2013, \apj, 770, 147

\bibitem[{{Childs} {et~al.}(2012){Childs}, {Brugger}, {Whitlock}, {Meredith},
  {Ahern}, {Bonnell}, {Miller}, {Weber}, {Harrison}, {Pugmire}, {Fogal},
  {Garth}, {Sanderson}, {Bethel}, {Durant}, {Camp}, {Favre}, {R\"{u}bel}, \&
  {Navr\'{a}til}}]{HPV:VisIt}
{Childs}, H., {Brugger}, E., {Whitlock}, B., {Meredith}, J., {Ahern}, S.,
  {Bonnell}, K., {Miller}, M., {Weber}, G.~H., {Harrison}, C., {Pugmire}, D.,
  {Fogal}, T., {Garth}, C., {Sanderson}, A., {Bethel}, E.~W., {Durant}, M.,
  {Camp}, D., {Favre}, J.~M., {R\"{u}bel}, O., \& {Navr\'{a}til}, P. 2012, in
  {High Performance Visualization---Enabling Extreme-Scale Scientific Insight},
  357--372

\bibitem[{{Daldorff} {et~al.}(2014){Daldorff}, {T{\'o}th}, {Gombosi},
  {Lapenta}, {Amaya}, {Markidis}, \& {Brackbill}}]{Daldorff2014}
{Daldorff}, L.~K.~S., {T{\'o}th}, G., {Gombosi}, T.~I., {Lapenta}, G., {Amaya},
  J., {Markidis}, S., \& {Brackbill}, J.~U. 2014, Journal of Computational
  Physics, 268, 236

\bibitem[{{Daughton} {et~al.}(2011){Daughton}, {Roytershteyn}, {Karimabadi},
  {Yin}, {Albright}, {Bergen}, \& {Bowers}}]{Daughton2011}
{Daughton}, W., {Roytershteyn}, V., {Karimabadi}, H., {Yin}, L., {Albright},
  B.~J., {Bergen}, B., \& {Bowers}, K.~J. 2011, Nature Physics, 7, 539

\bibitem[{{de Gouveia dal Pino} \& {Lazarian}(2005)}]{deGouveia2005}
{de Gouveia dal Pino}, E.~M. \& {Lazarian}, A. 2005, \aap, 441, 845

\bibitem[{{de Gouveia Dal Pino} {et~al.}(2010){de Gouveia Dal Pino},
  {Piovezan}, \& {Kadowaki}}]{deGouveia2010}
{de Gouveia Dal Pino}, E.~M., {Piovezan}, P.~P., \& {Kadowaki}, L.~H.~S. 2010,
  \aap, 518, A5

\bibitem[{{Del Santo} {et~al.}(2013){Del Santo}, {Malzac}, {Belmont},
  {Bouchet}, \& {De Cesare}}]{DelSanto2013}
{Del Santo}, M., {Malzac}, J., {Belmont}, R., {Bouchet}, L., \& {De Cesare}, G.
  2013, \mnras, 430, 209

\bibitem[{{Dexter} {et~al.}(2014){Dexter}, {McKinney}, {Markoff}, \&
  {Tchekhovskoy}}]{Dexter2013}
{Dexter}, J., {McKinney}, J.~C., {Markoff}, S., \& {Tchekhovskoy}, A. 2014,
  \mnras, 440, 2185

\bibitem[{{Di Matteo}(1998)}]{Matteo1998}
{Di Matteo}, T. 1998, \mnras, 299, L15

\bibitem[{{Di Matteo} {et~al.}(1997){Di Matteo}, {Blackman}, \&
  {Fabian}}]{Matteo1997}
{Di Matteo}, T., {Blackman}, E.~G., \& {Fabian}, A.~C. 1997, \mnras, 291, L23

\bibitem[{{D{\'{\i}}az} {et~al.}(2013){D{\'{\i}}az}, {Miller-Jones},
  {Migliari}, {Broderick}, \& {Tzioumis}}]{Trigo2013}
{D{\'{\i}}az}, Mar{\'{\i}}a, T., {Miller-Jones}, J.~C.~A., {Migliari}, S.,
  {Broderick}, J.~W., \& {Tzioumis}, T. 2013, \nat, 504, 260

\bibitem[{{Drake} {et~al.}(2010){Drake}, {Opher}, {Swisdak}, \&
  {Chamoun}}]{Drake2010}
{Drake}, J.~F., {Opher}, M., {Swisdak}, M., \& {Chamoun}, J.~N. 2010, \apj,
  709, 963

\bibitem[{Drake {et~al.}(2005)Drake, Shay, Thongthai, \& Swisdak}]{Drake2005}
Drake, J.~F., Shay, M.~A., Thongthai, W., \& Swisdak, M. 2005, Phys. Rev.
  Lett., 94, 095001

\bibitem[{{Drake} {et~al.}(2006){Drake}, {Swisdak}, {Che}, \&
  {Shay}}]{Drake2006}
{Drake}, J.~F., {Swisdak}, M., {Che}, H., \& {Shay}, M.~A. 2006, \nat, 443, 553

\bibitem[{{Drenkhahn} \& {Spruit}(2002)}]{Drenkhahn2002}
{Drenkhahn}, G. \& {Spruit}, H.~C. 2002, \aap, 391, 1141

\bibitem[{{Drury}(2012)}]{Drury2012}
{Drury}, L.~O. 2012, \mnras, 2661

\bibitem[{{Eastwood} {et~al.}(2013){Eastwood}, {Phan}, {{\O}ieroset}, {Shay},
  {Malakit}, {Swisdak}, {Drake}, \& {Masters}}]{Eastwood2013}
{Eastwood}, J.~P., {Phan}, T.~D., {{\O}ieroset}, M., {Shay}, M.~A., {Malakit},
  K., {Swisdak}, M., {Drake}, J.~F., \& {Masters}, A. 2013, Plasma Physics and
  Controlled Fusion, 55, 124001

\bibitem[{Egedal {et~al.}(2009)Egedal, Daughton, Drake, Katz, \&
  Lê}]{Egedal2009}
Egedal, J., Daughton, W., Drake, J.~F., Katz, N., \& Lê, A. 2009, Physics of
  Plasmas (1994-present), 16,

\bibitem[{{Egedal} {et~al.}(2012){Egedal}, {Daughton}, \& {Le}}]{Egedal2012}
{Egedal}, J., {Daughton}, W., \& {Le}, A. 2012, Nature Physics, 8, 321

\bibitem[{{Giannios}(2010)}]{Giannios2010}
{Giannios}, D. 2010, \mnras, 408, L46

\bibitem[{{Giannios} {et~al.}(2009){Giannios}, {Uzdensky}, \&
  {Begelman}}]{Giannios2009}
{Giannios}, D., {Uzdensky}, D.~A., \& {Begelman}, M.~C. 2009, \mnras, 395, L29

\bibitem[{{Giroletti} {et~al.}(2004){Giroletti}, {Giovannini}, {Feretti},
  {Cotton}, {Edwards}, {Lara}, {Marscher}, {Mattox}, {Piner}, \&
  {Venturi}}]{Giroletti2004}
{Giroletti}, M., {Giovannini}, G., {Feretti}, L., {Cotton}, W.~D., {Edwards},
  P.~G., {Lara}, L., {Marscher}, A.~P., {Mattox}, J.~R., {Piner}, B.~G., \&
  {Venturi}, T. 2004, \apj, 600, 127

\bibitem[{{Goodman} \& {Uzdensky}(2008)}]{Goodman2008}
{Goodman}, J. \& {Uzdensky}, D. 2008, \apj, 688, 555

\bibitem[{{Gould} \& {Schr{\'e}der}(1967)}]{Gould1967}
{Gould}, R.~J. \& {Schr{\'e}der}, G.~P. 1967, Physical Review, 155, 1404

\bibitem[{{Graf von der Pahlen} \& {Tsiklauri}(2014)}]{Pahlen2013}
{Graf von der Pahlen}, J. \& {Tsiklauri}, D. 2014, Physics of Plasmas, 21,
  012901

\bibitem[{{Higashimori} \& {Hoshino}(2012)}]{Higashimori2012}
{Higashimori}, K. \& {Hoshino}, M. 2012, Journal of Geophysical Research (Space
  Physics), 117, 1220

\bibitem[{{IceCube Collaboration}(2013)}]{IceCube2013}
{IceCube Collaboration}. 2013, Science, 342

\bibitem[{{Jaroschek} {et~al.}(2008){Jaroschek}, {Hoshino}, {Lesch}, \&
  {Treumann}}]{Jaroschek2008}
{Jaroschek}, C.~H., {Hoshino}, M., {Lesch}, H., \& {Treumann}, R.~A. 2008,
  Advances in Space Research, 41, 481

\bibitem[{{Jiang} {et~al.}(2014){Jiang}, {Stone}, \& {Davis}}]{Jiang2014}
{Jiang}, Y.-F., {Stone}, J.~M., \& {Davis}, S.~W. 2014, \apj, 784, 169

\bibitem[{{Kagan} {et~al.}(2013){Kagan}, {Milosavljevi{\'c}}, \&
  {Spitkovsky}}]{Kagan2012}
{Kagan}, D., {Milosavljevi{\'c}}, M., \& {Spitkovsky}, A. 2013, \apj, 774, 41

\bibitem[{{Kato}(2013)}]{Kato2013}
{Kato}, T.~N. 2013, ArXiv e-prints

\bibitem[{Khotyaintsev {et~al.}(2006)Khotyaintsev, Vaivads, Retin\`o, Andr\'e,
  Owen, \& Nilsson}]{Khotyaintsev2005}
Khotyaintsev, Y.~V., Vaivads, A., Retin\`o, A., Andr\'e, M., Owen, C.~J., \&
  Nilsson, H. 2006, Phys. Rev. Lett., 97, 205003

\bibitem[{{Kirk} \& {Skj{\ae}raasen}(2003)}]{Kirk2003}
{Kirk}, J.~G. \& {Skj{\ae}raasen}, O. 2003, \apj, 591, 366

\bibitem[{{Klimas} {et~al.}(2010){Klimas}, {Hesse}, {Zenitani}, \&
  {Kuznetsova}}]{Klimas2010}
{Klimas}, A., {Hesse}, M., {Zenitani}, S., \& {Kuznetsova}, M. 2010, Physics of
  Plasmas, 17, 112904

\bibitem[{{Kotani} {et~al.}(1994){Kotani}, {Kawai}, {Aoki}, {Doty}, {Matsuoka},
  {Mitsuda}, {Nagase}, {Ricker}, \& {White}}]{Kotani1994}
{Kotani}, T., {Kawai}, N., {Aoki}, T., {Doty}, J., {Matsuoka}, M., {Mitsuda},
  K., {Nagase}, F., {Ricker}, G., \& {White}, N.~E. 1994, \pasj, 46, L147

\bibitem[{{Kowal} {et~al.}(2011){Kowal}, {de Gouveia Dal Pino}, \&
  {Lazarian}}]{Kowal2011}
{Kowal}, G., {de Gouveia Dal Pino}, E.~M., \& {Lazarian}, A. 2011, \apj, 735,
  102

\bibitem[{{Kronberg} {et~al.}(2004){Kronberg}, {Colgate}, {Li}, \&
  {Dufton}}]{Kronberg2004}
{Kronberg}, P.~P., {Colgate}, S.~A., {Li}, H., \& {Dufton}, Q.~W. 2004, \apjl,
  604, L77

\bibitem[{{Lazar} {et~al.}(2009){Lazar}, {Nakar}, \& {Piran}}]{Lazar2009}
{Lazar}, A., {Nakar}, E., \& {Piran}, T. 2009, \apjl, 695, L10

\bibitem[{{Liu} {et~al.}(2014){Liu}, {Daughton}, {Karimabadi}, {Li}, \& {Peter
  Gary}}]{Liu2014}
{Liu}, Y.-H., {Daughton}, W., {Karimabadi}, H., {Li}, H., \& {Peter Gary}, S.
  2014, Physics of Plasmas, 21, 022113

\bibitem[{{Liu} {et~al.}(2012){Liu}, {Drake}, \& {Swisdak}}]{Liu2012}
{Liu}, Y.-H., {Drake}, J.~F., \& {Swisdak}, M. 2012, Physics of Plasmas, 19,
  022110

\bibitem[{{Lyutikov}(2006{\natexlab{a}})}]{Lyutikov2006c}
{Lyutikov}, M. 2006{\natexlab{a}}, \mnras, 369, L5

\bibitem[{{Lyutikov}(2006{\natexlab{b}})}]{Lyutikov2006b}
---. 2006{\natexlab{b}}, \mnras, 367, 1594

\bibitem[{{Markidis} {et~al.}(2013){Markidis}, {Henri}, {Lapenta}, {Divin},
  {Goldman}, {Newman}, \& {Laure}}]{Markidis2013}
{Markidis}, S., {Henri}, P., {Lapenta}, G., {Divin}, A., {Goldman}, M.,
  {Newman}, D., \& {Laure}, E. 2013, Physics of Plasmas, 20, 082105

\bibitem[{{May} {et~al.}(2014){May}, {Tonge}, {Mori}, {Fi{\'u}za}, {Fonseca},
  {Silva}, \& {Ren}}]{May2014}
{May}, J., {Tonge}, J., {Mori}, W.~B., {Fi{\'u}za}, F., {Fonseca}, R.~A.,
  {Silva}, L.~O., \& {Ren}, C. 2014, ArXiv e-prints

\bibitem[{{McKinney} {et~al.}(2012){McKinney}, {Tchekhovskoy}, \&
  {Blandford}}]{McKinney2012}
{McKinney}, J.~C., {Tchekhovskoy}, A., \& {Blandford}, R.~D. 2012, \mnras, 423,
  3083

\bibitem[{{McKinney} \& {Uzdensky}(2012)}]{McKinney2012b}
{McKinney}, J.~C. \& {Uzdensky}, D.~A. 2012, \mnras, 419, 573

\bibitem[{{Melzani} {et~al.}(2014{\natexlab{a}}){Melzani}, {Walder}, {Folini},
  \& {Winisdoerffer}}]{Melzani2013b}
{Melzani}, M., {Walder}, R., {Folini}, D., \& {Winisdoerffer}, C.
  2014{\natexlab{a}}, International Journal of Modern Physics Conference
  Series, 28, 60194

\bibitem[{{Melzani} {et~al.}(2014{\natexlab{b}}){Melzani}, {Walder}, {Folini},
  {Winisdoerffer}, \& {Favre}}]{Melzani2014a}
{Melzani}, M., {Walder}, R., {Folini}, D., {Winisdoerffer}, C., \& {Favre},
  J.~M. 2014{\natexlab{b}}, submitted to A\&A, ArXiv 1404.7366

\bibitem[{{Melzani} {et~al.}(2013){Melzani}, {Winisdoerffer}, {Walder},
  {Folini}, {Favre}, {Krastanov}, \& {Messmer}}]{Melzani2013}
{Melzani}, M., {Winisdoerffer}, C., {Walder}, R., {Folini}, D., {Favre}, J.~M.,
  {Krastanov}, S., \& {Messmer}, P. 2013, \aap, 558, A133

\bibitem[{{Merloni} \& {Fabian}(2001)}]{Merloni2001}
{Merloni}, A. \& {Fabian}, A.~C. 2001, \mnras, 321, 549

\bibitem[{{Meyer} {et~al.}(2010){Meyer}, {Horns}, \& {Zechlin}}]{Meyer2010}
{Meyer}, M., {Horns}, D., \& {Zechlin}, H.-S. 2010, \aap, 523, A2

\bibitem[{{Ohtani} \& {Horiuchi}(2009)}]{Ohtani2009}
{Ohtani}, H. \& {Horiuchi}, R. 2009, Plasma and Fusion Research, 4, 24

\bibitem[{Olshevsky {et~al.}(2013)Olshevsky, Lapenta, \&
  Markidis}]{Olshevsky2013}
Olshevsky, V., Lapenta, G., \& Markidis, S. 2013, Phys. Rev. Lett., 111, 045002

\bibitem[{{Pei} {et~al.}(2001){Pei}, {Horiuchi}, \& {Sato}}]{Pei2001}
{Pei}, W., {Horiuchi}, R., \& {Sato}, T. 2001, Physics of Plasmas, 8, 3251

\bibitem[{{P{\'e}tri} \& {Lyubarsky}(2007)}]{Petri2007b}
{P{\'e}tri}, J. \& {Lyubarsky}, Y. 2007, \aap, 473, 683

\bibitem[{{Petropoulou} {et~al.}(2014){Petropoulou}, {Giannios}, \&
  {Dimitrakoudis}}]{Petropoulou2014}
{Petropoulou}, M., {Giannios}, D., \& {Dimitrakoudis}, S. 2014, ArXiv e-prints

\bibitem[{{Pontin}(2011)}]{Pontin2011}
{Pontin}, D.~I. 2011, Advances in Space Research, 47, 1508

\bibitem[{{Poutanen} \& {Veledina}(2014)}]{Poutanen2014}
{Poutanen}, J. \& {Veledina}, A. 2014, \ssr

\bibitem[{{Pritchett}(2005{\natexlab{a}})}]{Pritchett2005b}
{Pritchett}, P.~L. 2005{\natexlab{a}}, Journal of Geophysical Research (Space
  Physics), 110, 5209

\bibitem[{{Pritchett}(2005{\natexlab{b}})}]{Pritchett2005}
---. 2005{\natexlab{b}}, Journal of Geophysical Research (Space Physics), 110,
  10213

\bibitem[{{Reis} \& {Miller}(2013)}]{Reis2013}
{Reis}, R.~C. \& {Miller}, J.~M. 2013, \apjl, 769, L7

\bibitem[{{Romanova} \& {Lovelace}(1992)}]{Romanova1992}
{Romanova}, M.~M. \& {Lovelace}, R.~V.~E. 1992, \aap, 262, 26

\bibitem[{{Romero} {et~al.}(2014){Romero}, {Vieyro}, \& {Chaty}}]{Romero2014}
{Romero}, G.~E., {Vieyro}, F.~L., \& {Chaty}, S. 2014, \aap, 562, L7

\bibitem[{{Rybicki} \& {Lightman}(1979)}]{Rybicki1979}
{Rybicki}, G.~B. \& {Lightman}, A.~P. 1979, {Radiative processes in
  astrophysics}, ed. {Rybicki, G.~B.~\& Lightman, A.~P.}

\bibitem[{{Schwartz} {et~al.}(2006){Schwartz}, {Marshall}, {Lovell}, {Murphy},
  {Bicknell}, {Birkinshaw}, {Gelbord}, {Georganopoulos}, {Godfrey}, {Jauncey},
  {Perlman}, \& {Worrall}}]{Schwartz2006}
{Schwartz}, D.~A., {Marshall}, H.~L., {Lovell}, J.~E.~J., {Murphy}, D.~W.,
  {Bicknell}, G.~V., {Birkinshaw}, M., {Gelbord}, J., {Georganopoulos}, M.,
  {Godfrey}, L., {Jauncey}, D.~L., {Perlman}, E.~S., \& {Worrall}, D.~M. 2006,
  \apj, 640, 592

\bibitem[{{Sironi} \& {Spitkovsky}(2011{\natexlab{a}})}]{Sironi2011b}
{Sironi}, L. \& {Spitkovsky}, A. 2011{\natexlab{a}}, \apj, 741, 39

\bibitem[{{Sironi} \& {Spitkovsky}(2011{\natexlab{b}})}]{Sironi2011}
---. 2011{\natexlab{b}}, \apj, 726, 75

\bibitem[{{Sironi} \& {Spitkovsky}(2014)}]{Sironi2014}
---. 2014, \apjl, 783, L21

\bibitem[{{Uzdensky} {et~al.}(2011){Uzdensky}, {Cerutti}, \&
  {Begelman}}]{Uzdensky2011}
{Uzdensky}, D.~A., {Cerutti}, B., \& {Begelman}, M.~C. 2011, \apjl, 737, L40

\bibitem[{{Vapirev} {et~al.}(2013){Vapirev}, {Lapenta}, {Divin}, {Markidis},
  {Henri}, {Goldman}, \& {Newman}}]{Vapirev2013}
{Vapirev}, A.~E., {Lapenta}, G., {Divin}, A., {Markidis}, S., {Henri}, P.,
  {Goldman}, M., \& {Newman}, D. 2013, Journal of Geophysical Research (Space
  Physics), 118, 1435

\bibitem[{{Walder} {et~al.}(2014){Walder}, {Melzani}, {Folini},
  {Winisdoerffer}, \& {Favre}}]{Walder2014}
{Walder}, R., {Melzani}, M., {Folini}, D., {Winisdoerffer}, C., \& {Favre},
  J.~M. 2014, ASTRONUM 2013, Conf. Proceedings, in press, astro-ph:1405.0600

\bibitem[{{Zdziarski} {et~al.}(2014){Zdziarski}, {Pjanka}, {Sikora}, \&
  {Stawarz}}]{Zdziarski2014}
{Zdziarski}, A.~A., {Pjanka}, P., {Sikora}, M., \& {Stawarz}, L. 2014, ArXiv
  e-prints

\bibitem[{{Zenitani} \& {Hoshino}(2001)}]{Zenitani2001}
{Zenitani}, S. \& {Hoshino}, M. 2001, \apjl, 562, L63

\bibitem[{{Zenitani} \& {Hoshino}(2007)}]{Zenitani2007}
---. 2007, \apj, 670, 702

\end{thebibliography}
\addcontentsline{toc}{section}{References}

\appendix

\section{The importance of radiative braking, Compton drag, and pair creations}
\label{sec:appendix_radiation}

Two relevant issues are the importance of radiative braking and of pair creation.
We first investigate radiative braking of electrons, and then study the opacity of high energy photons 
to $\gamma\gamma$ annihilations.

\subsubsection*{Electron braking by emission of radiation or by Compton drag}
Electrons lose energy by emitting photons when being scattered by magnetic fields 
(synchrotron-like radiation) or by Coulomb collisions (Bremsstrahlung-like radiation), 
or when colliding with photons (inverse-Compton events).
For the synchrotron component, the energy $\delta E_\mathrm{sync}$ lost by an electron 
of Lorentz factor $\gamma$ and velocity $\beta c$, gyrating in a magnetic field $B$, 
averaged over pitch angles, and during one cyclotron orbit, is \citep{Rybicki1979}:
\begin{equation}\label{equ:sync_emmision_braking}
 \frac{\delta E_\mathrm{sync}}{\gamma m_\lec c^2} = \frac{8\pi}{9} \beta^2\gamma^2 \frac{r_0\wce}{c} = \frac{B}{1.4\times10^{11}\mathrm{G}} \, \left(\frac{\gamma}{100}\right)^2,
\end{equation}
with $r_0$ the classical electron radius and $\wce=eB/m_\lec$.
On the other hand, electron cooling by Coulomb collisions can be evaluated 
via the thermal Bremsstrahlung emission formula, giving
an energy $\delta E_\mathrm{Brem}$ lost during one plasma period $\wpe^{-1}$:
\begin{equation}\label{equ:Brem_emmision_braking}
 \frac{\delta E_\mathrm{Brem}}{m_\lec c^2} = \left(\frac{T_\lec}{m_\lec c^2}\right)^{1/2} \, \left(\frac{n_\lec}{5\times10^{12}\,\mathrm{cm^{-3}}}\right)^{3/2}.
\end{equation}
It shows that synchrotron braking is not relevant for the objects of Table~\ref{tab:real_parameters},
except for pulsar wind nebulae and very high Lorentz factor electrons,
while braking by Bremsstrahlung emission is significant for 
reconnection in microquasar magnetospheres close to the black hole. 

The last braking mechanism is inverse-Compton scattering of ambient photons by electrons. 
The energy $\delta E$ lost by an electron during one plasma period $\wpe^{-1}$ is at most 
\begin{equation}\label{equ:IC_loss}
\begin{aligned}
 \frac{\delta E_\mathrm{IC}}{\gamma m_\lec c^2} &= \frac{4}{3} \frac{\sigma_\mathrm{T} c \beta^2\gamma}{\wpe} \frac{U_\mathrm{ph}}{m_\lec c^2} \\
  &= \frac{\gamma\beta^2}{100} \, \left(\frac{1\,\mathrm{cm^{-3}}}{n_\lec}\right)^{1/2} \, \frac{U_\mathrm{ph}}{1.7\times10^{10}\mathrm{erg/cm^3}},
\end{aligned}
\end{equation}
where $U_\mathrm{ph}$ is the photon energy density and $\sigma_\mathrm{T}$ is Thomson cross section.
For a blackbody radiation, the energy density is given by 
$U_\mathrm{ph} = (\pi^2/15) \, T_\mathrm{ph}^4 / (\hbar c)^3$. It reaches the density $1.7\times10^{10}\mathrm{erg/cm^3}$
for $T_\mathrm{ph} = 1.2\times10^6\mathrm{K}$. 
Below this temperature, electrons do not significantly lose energy by Compton drag, while above they do.

However, it should be noted that a blackbody radiation at $T_\mathrm{ph}$ 
requires the thermalization of the photons produced by the hot electrons, 
a fact impossible to achieve on an inertial length scale $d_\lec$ given 
that the mean-free-path to Compton scattering is
\begin{equation}\label{equ:opacity_Compton}
 \frac{l_\mathrm{Compt}}{d_\lec} = \frac{1}{d_\lec \sigma_\mathrm{T} n_\lec} = \left(\frac{9\times10^{36}\,\mathrm{cm^{-3}}}{n_\lec}\right)^{1/2},
\end{equation}
and would thus implies densities $n_\lec \sim 10^{36}\,\mathrm{cm^{-3}}$.
The radiation produced by the hot electrons in the reconnection region 
consequently escapes from this region before being thermalized.
If we assume that the photons are produced over a volume $(ad_\lec)^3$, with $a$ a geometrical factor,
and that the emissivity is $\dif W_\mathrm{ph}/\dif t \dif V$, then 
the energy density of the gas of photons is 
$U_\mathrm{ph} \sim ad_\lec / c \times \dif W_\mathrm{ph}/\dif t \dif V$.
Expressions for $U_\mathrm{ph}$ should thus be obtained for synchrotron and Bremsstrahlung radiations.

Concerning synchrotron radiation, the total power emitted by an electron is 
$P_\mathrm{emit,sync}=(4/9)r_0^2 c \beta^2 \gamma^2B^2$,
and the emissivity is $\sim n_\lec P_\mathrm{emit,sync}$, so that
\begin{equation}
\begin{aligned}
 U_\mathrm{ph,sync} = &9\times10^{-26}\,\mathrm{erg/cm^3} \\
  & \times a\, \left(\frac{B}{1\,\mathrm{G}}\right)^2 \, \left(\frac{n_\lec}{1\,\mathrm{cm^{-3}}}\right)^{1/2} \, \left(\frac{\gamma\beta}{100}\right)^2.
\end{aligned}
\end{equation}
This expression can be inserted into Eq.~\ref{equ:IC_loss}, to yield:
\begin{equation}\label{equ:sync_compton_drag}
 \frac{\delta E_\mathrm{IC,sync}}{\gamma m_\lec c^2} = 5.3\times10^{-36} \times a\, \left(\frac{B}{1\,\mathrm{G}}\right)^2 \, \beta\left(\frac{\gamma\beta}{100}\right)^3,
\end{equation}
which is always well below unity for objects of Table~\ref{tab:real_parameters}.

As for Bremsstrahlung radiation, we use the emissivity for a thermalized plasma at temperature $T_\lec$, to find
\begin{equation}
\begin{aligned}
 U_\mathrm{ph,Brem} = & 2.5\times10^{-28}\,\mathrm{erg/cm^3} \\
 & \times a\, \left(\frac{T_\lec}{10^8\,\mathrm{K}}\right)^{1/2} \, \left(\frac{n_\lec}{1\,\mathrm{cm^{-3}}}\right)^{3/2}.
\end{aligned}
\end{equation}
Once inserted into Eq.~\ref{equ:IC_loss}, it yields:
\begin{equation}\label{equ:Brem_compton_drag}
 \frac{\delta E_\mathrm{IC,Brem}}{\gamma m_\lec c^2} = 1.5\times10^{-38} \times a \, \left(\frac{T_\lec}{10^8\,\mathrm{K}}\right)^{1/2} \, \left(\frac{n_\lec}{1\,\mathrm{cm^{-3}}}\right) \, \frac{\gamma\beta^2}{100},
\end{equation}
which is again always well below unity for objects of Table~\ref{tab:real_parameters}.

\subsubsection*{High-energy photons and opacity to $\gamma\gamma$-annihilation}
The photons emitted by the accelerated electrons 
(of Lorentz factor $\gamma$, velocity $\beta c$) can be due either to synchrotron or to Bremsstrahlung radiation.
In the synchrotron case, photon energies can reach 
\begin{equation}\label{equ:hnu_sync}
\begin{aligned}
 h\nu_\mathrm{sync} &= \frac{3\gamma^2\wce}{2\pi\beta} \\
                    &= 3.5\times10^{-4}\,\mathrm{eV} \, \frac{B}{1\,\mathrm{G}} \, \left(\frac{\gamma}{100}\right)^2 \\
                    &= 6.8\times10^{-10}m_\lec c^2 \, \frac{B}{1\,\mathrm{G}} \, \left(\frac{\gamma}{100}\right)^2.
\end{aligned}
\end{equation}
Also, a thermal Bremsstrahlung spectra from electrons at temperature $T_\lec$ cuts-off above 
\begin{equation}\label{equ:hnu_Brem}
 h\nu_\mathrm{therm} = m_\lec c^2 \, \frac{T_\lec}{6\times10^9\,\mathrm{K}}.
\end{equation}
High-energy photons can annihilate with lower energy photons of energy $\epsilon_0$
only if they have an energy above $m_\lec^2 c^4/\epsilon_0$ \citep{Gould1967}, which requires at least a
high-energy photon above $0.5\mathrm{MeV}$.
Synchrotron radiation can produce such photons in pulsar wind nebulae,
and thermal Bremsstrahlung can do so in microquasars and GRBs.

Also, high-energy photons can be produced by inverse-Compton collisions between ambient photons of 
energy $\epsilon_0$ and high-energy electrons of Lorentz factor $\gamma$. 
The outcome of such a collision is a high-energy photon of energy up to $\gamma^2\epsilon_0$, 
so that $\gamma > m_\lec c^2 / \epsilon_0$ is needed to produce pairs.

It is then interesting to compute the mean-free-path $l_{\gamma\gamma}$ of such high-energy photons. 
Their annihilation creates pairs, and will affect the reconnection dynamics only if they are created near the 
reconnection site. The relevant quantity is thus $l_{\gamma\gamma} / d_\lec$, with $d_\lec$ the electron inertial length.
We only seek an order of magnitude estimate. From \citet{Gould1967}, we can approximate the optical depth $\tau_{\gamma\gamma}$
for a high-energy photon (energy $E$) traveling a length $l$ through a gas of lower energy photons (with a typical energy $\epsilon_0$, 
of number density $n_\mathrm{ph}$) as
\begin{equation}
 \tau_{\gamma\gamma} \sim l\, \pi r_0^2 \, n_\mathrm{ph} \, f(m_\lec^2 c^4 / \{E\epsilon_0\}),
\end{equation}
where $r_0$ is the classical electron radius, 
and $f$ is a function depending on the exact gas photon distribution. 
Generally, $f$ is maximal and equal to $\sim 1$ for $E = \epsilon_0$. 
For example, if the gas of photons is a blackbody at temperature $T_\mathrm{ph}$, then 
$\epsilon_0 = T_\mathrm{ph}$, $n_\mathrm{ph} = 2\zeta(3)/\pi^2\, (T_\mathrm{ph}/(\hbar c))^3$, 
with $\zeta(3)\sim 1.202$,
and $f$ is maximal for $E = T_\mathrm{ph}$ with a value $1/2\zeta(3)$.
The mean-free-path is defined such that $\tau_{\gamma\gamma}=1$.
In the blackbody case, we have
\begin{equation}\label{equ:gammagamma_annihilation_CN}
 \frac{l_{\gamma\gamma,\mathrm{BB}}}{d_\lec} \sim \frac{1}{\pi r_0^2 d_\lec n_\mathrm{ph}} = \left(\frac{1\,\mathrm{cm^{-3}}}{n_\lec}\right)^{1/2} 
             \, \left(\frac{1.4\times10^6\,\mathrm{K}}{T_\mathrm{ph}}\right)^3.
\end{equation}
Consequently, if $T_\mathrm{ph} < 1.4\times10^6\,\mathrm{K}$, then the $\gamma\gamma$ annihilations occur well outside the 
reconnection region and does not affect the process; while for larger photon temperatures 
the annihilation occurs after a free flight of less than an inertial length,
i.e., inside the reconnection region.
For an unspecified photon number density, Eq.~\ref{equ:gammagamma_annihilation_CN} can be written
\begin{equation}\label{equ:gammagamma_annihilation_any}
 \frac{l_{\gamma\gamma}}{d_\lec} \sim \frac{1}{\pi r_0^2 d_\lec n_\mathrm{ph}} = \left(\frac{1\,\mathrm{cm^{-3}}}{n_\lec}\right)^{1/2} 
              \, \frac{7\times10^{19}\,\mathrm{cm^{-3}}}{n_\mathrm{ph}}.
\end{equation}
However, as previously noted, a blackbody radiation at $T_\mathrm{ph}$ requires the thermalization 
of the photons by the hot electrons, 
a fact impossible to achieve if these electrons are confined to the reconnection region 
on inertial length scales (Eq.~\ref{equ:opacity_Compton}).
Again, the radiation number density should be estimated as 
$n_\mathrm{ph} \sim ad_\lec / c \times \dif n_\mathrm{ph}/\dif t$,
with $\dif n_\mathrm{ph}/\dif t$ the production rate of photons and $(ad_\lec)^3$ the volume of the emission region.
We then evaluate $n_\mathrm{ph}$ for synchrotron and Bremsstrahlung radiation.

For synchrotron radiation, 
this can be roughly estimated by dividing the total power emitted by an electron by the characteristic energy 
$h\nu_\mathrm{sync}$ (Eq.~\ref{equ:hnu_sync}), and then multiplying by the electron number density.
After some manipulations, one arrives at $\dif n_\mathrm{ph}/\dif t = P_\mathrm{emit}/(h\nu_\mathrm{sync}) = 0.1\wce$, 
so that 
\begin{equation}
 n_\mathrm{ph,sync} = 0.1a \frac{\wce}{\wpe}\,n_\lec,
\end{equation}
with $\wpe$ the plasma pulsation associated to $d_\lec$.
We can rewrite $\wce/\wpe = (\sigma^\mathrm{cold}_\lec)^{1/2}$ with $\sigma^\mathrm{cold}_\lec = B^2 / (\mu_0 n_\lec m_\lec c^2)$.
Inserting $n_\mathrm{ph,sync}$ into Eq.~\ref{equ:gammagamma_annihilation_any}, we have
\begin{equation}\label{equ:gammagamma_annihilation_sync}
 \frac{l_{\gamma\gamma,\mathrm{sync}}}{d_\lec} \sim a^{-1}\,\left(\frac{1\,\mathrm{cm^{-3}}}{n_\lec}\right)^{3/2} 
      \, \left(\frac{3\times10^{10}}{\sigma^\mathrm{cold}_\lec}\right)^{1/2}.
\end{equation}
For radio lobes, radio emitting regions of extragalactic jets, or pulsar wind nebulae,
$l_{\gamma\gamma,\mathrm{sync}} \gg d_\lec$ holds, so that pairs form far away from the reconnection site; 
while for microquasar coronae close to the hole, for extragalactic jet $\gamma$-ray region,
for GRB jets, or for pulsar wind termination shocks, we have 
$l_{\gamma\gamma,\mathrm{sync}} \ll d_\lec$ and pairs form inside the reconnection region.

For Bremsstrahlung radiation, one can estimate the photon number density 
by dividing the Bremsstrahlung emissivity $P_\mathrm{emit}$
by the typical energy $T_\lec$, and multiplying by a photon escape length $ad_\lec$.
One finds 
\begin{equation}
 n_\mathrm{ph,Brem} = 1.8\times10^{-20}\,\mathrm{cm^{-3}} \, a\, \left(\frac{10^8\,\mathrm{K}}{T_\lec}\right)^{1/2} \, \left(\frac{n_\lec}{1\,\mathrm{cm^{-3}}}\right)^{3/2}.
\end{equation}
Inserting into Eq.~\ref{equ:gammagamma_annihilation_any}, we have
\begin{equation}\label{equ:gammagamma_annihilation_Brem}
 \frac{l_{\gamma\gamma,\mathrm{Brem}}}{d_\lec} \sim 4\times10^{39}\times a^{-1} \left(\frac{1\,\mathrm{cm^{-3}}}{n_\lec}\right)^{2} 
     \, \left(\frac{T_\lec}{10^8\,\mathrm{K}}\right)^{1/2}.
\end{equation}
We have $l_{\gamma\gamma,\mathrm{Brem}}\gg d_\lec$ for all objects of 
Table~\ref{tab:real_parameters}.

\end{document}